\documentclass[a4paper,fleqn,usenatbib]{mnras}

\usepackage{mathptmx}
\usepackage{txfonts}

\usepackage[T1]{fontenc}
\usepackage{ae,aecompl}

\usepackage{graphicx}	

\newcommand{\Msun}{M$_{\odot}$}
\newcommand{\kms}{km\,s$^{-1}$} 
\newcommand{\Omratio}{$\Omega / \Omega_{crit}$}
\newcommand{\halpha}{H${\alpha}$}

\title[On the eMSTO of NGC\,1850]{Dissecting the Extended Main Sequence Turn-off
of the Young Star Cluster NGC\,1850\thanks{Based on observations with the
NASA/ESA {\it Hubble Space Telescope}, obtained at the Space Telescope Science
Institute, which is operated by the Association of Universities for Research in
Astronomy, Inc., under NASA contract NAS5-26555}}

\author[M. Correnti et al.]{Matteo Correnti$^{1}$\thanks{E-mail: correnti@stsci.edu}, Paul Goudfrooij$^{1}$, Andrea Bellini$^{1}$, Jason S. Kalirai$^{1,2}$ 
\newauthor and Thomas H. Puzia$^{3}$\\
$^{1}$Space Telescope Science Institute, 3700 San Martin Drive, Baltimore, MD 21218, USA\\
$^{2}$Center for Astrophysical Sciences, 3400 N. Charles Street, John Hopkins University,
Baltimore, MD 21218, USA\\
$^{3}$Institute of Astrophysics, Pontificia Universidad Cat\'olica de Chile, Av.\ Vicu\~{n}a Mackenna 4860, Macul 
7820436, Santiago, Chile}
\usepackage{graphicx}

\pagerange{\pageref{firstpage}--\pageref{lastpage}} \pubyear{xxxx}

\begin{document}

\date{Accepted 2016 December 20. Received 2016 December 12; in original form 2016 November 07}

\maketitle

\label{firstpage}

\begin{abstract}
We use the Wide Field Camera 3 onboard the {\it Hubble Space Telescope} to
obtain deep, high-resolution photometry of the young ($\sim$ 100 Myr) star
cluster NGC\,1850 in the Large Magellanic Cloud. We analyze the cluster 
colour-magnitude diagram (CMD) and find that it hosts an extended main-sequence
turn-off (MSTO) and a double MS. We demonstrate that these features cannot be
due to photometric errors, field star contamination, or differential reddening.
From a comparison with theoretical models and Monte Carlo simulations, we show
that a coeval stellar population featuring a distribution of stellar rotation
rates can reproduce the MS split quite well. However, it \emph{cannot} reproduce
the observed MSTO region, which is significantly wider than the simulated ones.
Exploiting narrow-band \halpha\, imaging, we find that the MSTO hosts a
population of \halpha-emitting stars which are interpreted as rapidly rotating
Be-type stars. We explore the possibility that the discrepancy between the
observed MSTO morphology and that of the simulated simple stellar population
(SSP) is caused by the fraction of these objects that are highly reddened, but
we rule out this hypothesis. We demonstrate that the global CMD morphology is
well-reproduced by a combination of SSPs that cover an age range of $\sim$ 35
Myr {\em as well as} a wide variety of rotation rates. We derive the cluster
mass and escape velocity and use dynamical evolution models to predict their
evolution starting at an age of 10 Myr. We discuss these results and their
implications in the context of the extended MSTO phenomenon.
\end{abstract}


\section{Introduction}
\label{s:intro}
Deep colour-magnitude diagrams (CMDs) from images taken with the Advanced Camera
for Survey (ACS) and the Wide Field Camera 3 (WFC3) onboard the {\it Hubble
Space Telescope} ({\it HST}) revealed that several intermediate-age ($\sim$
1\,--2\, Gyr) star clusters in the Magellanic Clouds host extended main-sequence
turn-off (MSTO) regions \citep{mack+08a,
glat+08,milo+09,goud+09,goud+11b,goud+14,corr+14}, in some cases accompanied by
composite red clumps \citep{gira+09,rube+11}. The presence of these extended
MSTOs (hereafter eMSTOs) has been interpreted as due to stars that formed at
different times within the parent cluster, with age spreads of 150\,--500\, Myr
\citep{milo+09,gira+09,rube+10,rube+11,goud+11b,goud+14,kell+12,corr+14}. If so,
these objects might be the younger counterparts of the old  Galactic globular
clusters with multiple stellar populations \citep[e.g.,][]{conspe11}. An alternative scenario to explain the eMSTO phenomenon was originally proposed
by \citet{basdem09}, who suggested that this is due to the presence of a spread
in rotation velocity among turn-off stars \citep[for a detailed discussion,
see][and references therein]{gira+11,goud+14,nied+15,brahua15}. Finally,
\citet{yang+11} suggested that the eMSTOs and the dual clumps can be photometric
features produced by merged binary systems and interacting binaries with mass
transfer. 

Recently, eMSTOs have been detected also in young ($\sim$ 100\,--300\, Myr)
massive star clusters in the Large Magellanic Cloud (LMC). In particular,
\citet{milo+15a} and \citet[][hereafter C15]{corr+15} discovered the presence of
an eMSTO in the star cluster NGC\,1856, consistent with the presence of an age
spread of $\sim$ 80 Myr. Interestingly, the CMD of NGC\,1856 also shows a split
in the MS \citep{milo+15a}, which has been interpreted as an indication that the
cluster might host two coeval stellar populations with different rotational
velocities \citep{dant+15}. \citet{milo+16a} found that the CMD of the young
($\sim$ 100 Myr) star cluster NGC\,1755 shows a similar morphology, including 
an eMSTO and a MS split. They demonstrated that the bimodal MS is well
reproduced by a single stellar population with different rotation velocities,
whereas the fit between the observed eMSTO and models with different rotation is
not fully satisfactory. On the other hand, \citet{milo+13} showed that the
$\sim$ 300 Myr cluster NGC\,1844 presents only a split in the MS and not an
eMSTO. It is evident that these new discoveries in young clusters give us a new
opportunity to study the nature of the eMSTO phenomenon and to provide
additional constraints to discriminate among the different scenarios.    

\begin{figure*}
\centerline{
\includegraphics[height=8.2cm]{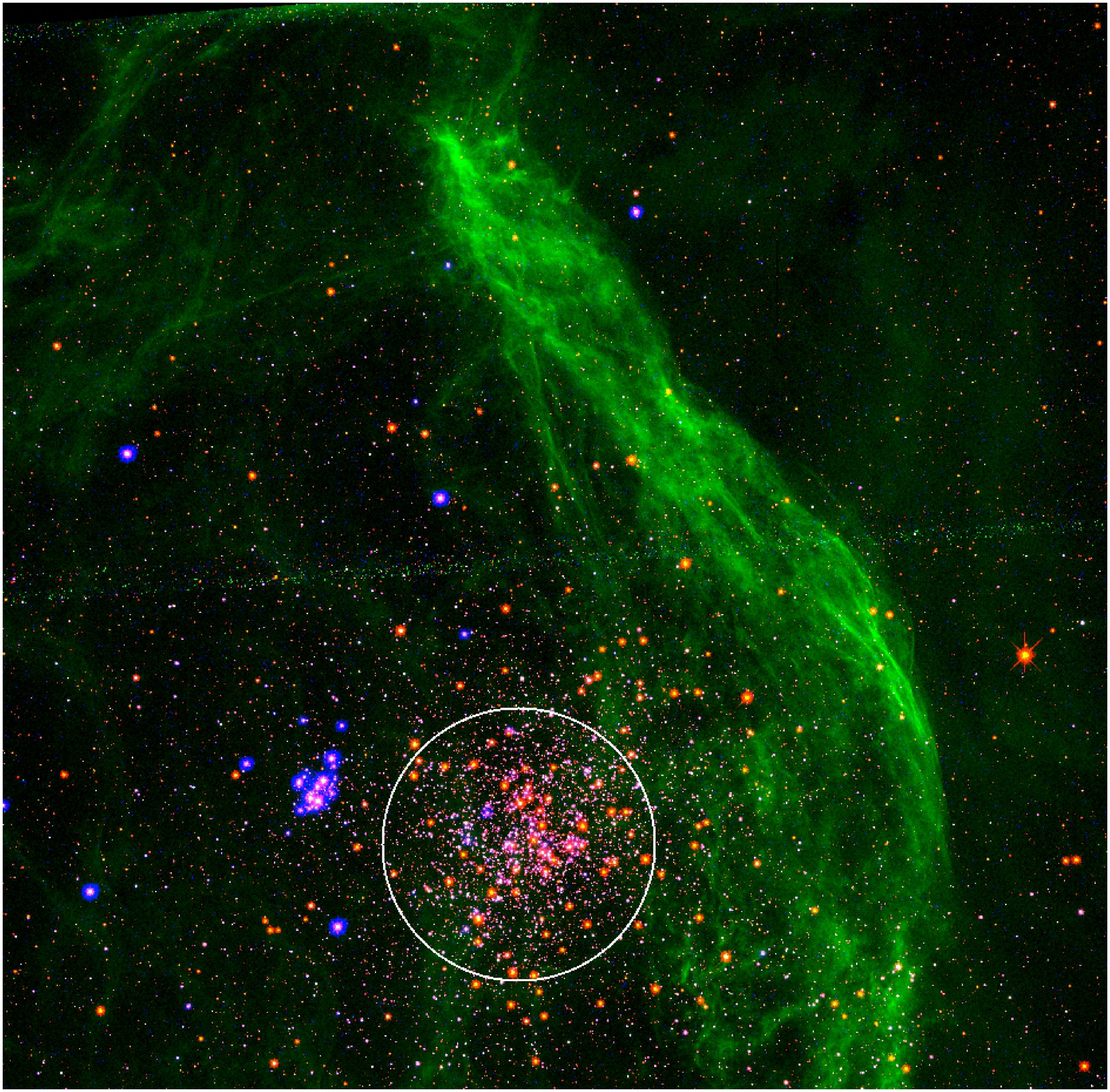}
\includegraphics[height=8.2cm]{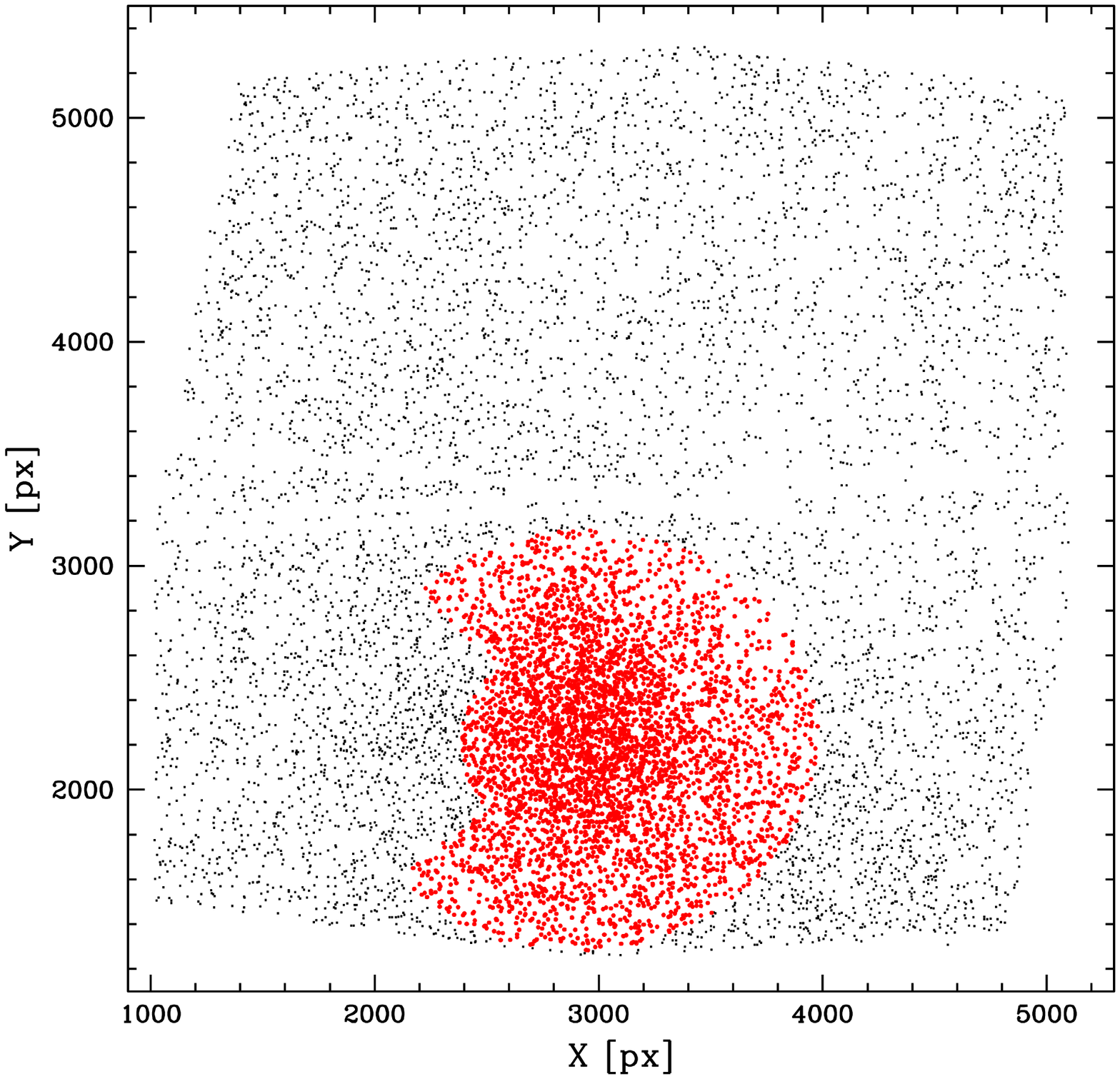}
}
\caption{Left panel: stacked trichromatic image of the WFC3/UVIS field (red
channel: {\em F814W} filter; green channel: {\em F656N} filter; blue channel:
{\em F275W} filter). The white circle marks a region of $\approx$ 20\arcsec\
which corresponds to the cluster effective radius $r_e$. The young cluster
NGC\,1850B is clearly visible on the left side of the cluster as well as the gas
filaments that stretch from the upper-left corner to the bottom-right corner,
covering a large portion of the image. Right panel: {\it XY} spatial
distribution for all the stars in the WFC3/UVIS field of view. Stars inside the
region considered as the cluster field in our analysis are reported as red dots.
The annulus segment on the left side has been excluded, in order to avoid
contamination by young stars belonging to NGC\,1850B.}  
\label{f:fov}
\end{figure*}

In this context, another interesting object is the young ($\sim$ 100 Myr) star
cluster NGC\,1850 in the LMC. \citet{bast+16} used {\it HST} WFC3 observations
with the {\em F336W} and {\em F438W} filters to show that the cluster hosts an
eMSTO which can be reproduced by an age spread of $\sim$ 40 Myr. However, they
suggest that this is the age range expected when the effect of rotation in MSTO
stars is interpreted in terms of an age spread, and thus concluded that no age
spread is present in the cluster. They also suggest that the presence of an age
spread should cause a ``bimodal'' distribution in the MSTO region, whereas the
observed distribution can be approximated by a single (possibly skewed) Gaussian
distribution\footnote[1]{This suggestion will be addressed in Sect.\,\ref{s:mc_sim}.}.  

Here, we present an analysis of new {\it HST} WFC3 photometry of NGC\,1850,
exploiting a different set of filters with respect to that used by 
\citet{bast+16}. Our data, thanks to a significantly larger colour baseline,
allow us to study thoroughly the CMD morphology and  to perform a detailed
analysis of the MSTO region. We compare the cluster CMD with theoretical models,
using isochrones with different ages and different rotation rates and with Monte
Carlo simulations in order to quantify to what extent the CMD can be reproduced
by a simple stellar population (SSP) with a distribution of rotation rates. We
derive the cluster mass and escape velocity to test whether the cluster has (or
has had) the right properties to retain mass-loss material that can be (or could
have been) used to form a second generation of stars. This study allows us to
have a clearer picture of the cluster's star formation history and in particular
on the nature of the observed eMSTO region. 

The remainder of the paper is organized as follows: observations and data
reduction are presented in Section\,2. In Section\,3, we discuss the observed
CMD, whereas in Section\,4 we compare the observed CMD with theoretical models,
using isochrones with different ages and different rotation rates. In
Section\,5, we compare the observed CMD with Monte Carlo simulations, both using
one SSP with different rotation rates and a combination of four SSPs with
different ages and different rotation rates; we obtain the pseudo-colour
distributions for the observed and simulated CMDs and compare them. In
Section\,6, we present the analysis of narrow-band \halpha\, imaging. In
Section\,7, we investigate how the CMD morphology changes as a function of the
distance from the cluster center,  while the physical and dynamical properties
of the cluster are presented in Section\,8. Finally, in Section\,9, we discuss
our results and their implications on the interpretation of the eMSTO
phenomenon.  

\section{Observations and Data Reduction}
\label{s:data}
NGC\,1850 was observed with {\it HST\/} on 2015 October 22 using the UVIS
channel of the WFC3 as part of the {\it HST} program 14174 (PI: P. Goudfrooij).
The cluster was centered on one of the two CCD chips of the WFC3/UVIS camera, to
avoid the loss of its central region due to the CCD chip gap and to minimize the
impact of different photometric zeropoints for the two chips. The cluster was
observed using three filters, namely {\em F275W}, {\em F656N}, and {\em F814W}.
Two long exposures were taken in the {\em F275W} and {\em F814W} filters, and
three in the {\em F656N} filter. The total exposure times for each filter are
1570 s ({\em F275W}), 2875 s ({\em F656N}), and 790 s ({\em F814W}). In
addition, we took two short exposures in the {\em F275W} and {\em F814W} filters
(150 s and 10 s, respectively), to avoid saturation of the brightest stars. The
two or three long exposures were spatially offset from each other by 0\farcs18
and -2\farcs401 in a direction +85\fdg76 with respect to the positive X- and
Y-axis of the detector. This was done to move across the gap between the two
WFC3/UVIS CCD chips, as well to simplify the identification and removal of hot
pixels. To investigate the \halpha-emitting stars in the cluster, we
complemented our dataset with archival observations with the filter {\em F467M}
(GO-11925, PI: S. E. Deustua). The main purpose of this filter in this context
is to define the continuum level at the wavelength of \halpha\, along with the
{\em F814W} filter. Program GO-11925 is a calibration program to test the
linearity of WFC3/UVIS, hence images in this filter were taken with various
exposure times; for our purpose, only the two long exposures were useful (300 s
and 500 s, respectively). In addition to the WFC3/UVIS observations, we used the
Wide Field Camera (WFC) of ACS in parallel to obtain images $\sim$ 6\arcmin\
from the cluster center using the {\em F555W}, {\em F658N}, and {\em F814W}
filters. These observations provide a clean picture of the stellar content, star
formation history, and H${\alpha}$-emitting  stars in the underlying LMC field. 

To reduce the images, we use the following method. Briefly, we start from the
{\it flt} files provided by the {\it HST} pipeline, which constitute the
bias-corrected, dark-subtracted, and flat-fielded images. The {\it flt} files
are first corrected for charge transfer inefficiency effects using the dedicated
CTE correction
software\footnote[2]{http://www.stsci.edu/hst/wfc3/tools/cte\_tools}. Star
positions and fluxes are measured on the CTE-corrected {\it flc} files using the
publicly-available FORTRAN program {\em img2xym\_wfc3uv} with the available
library of spatially variable point-spread functions (PSFs, in an array of $7
\times 8$ PSFs) for each filter. We derive an additional $5 \times 5$ array of
perturbation PSFs for each exposure, to account for telescope breathing effects
\citep{bell+14}, and we combine it with the library PSFs to fit stellar
profiles. Stellar positions are corrected for geometric distortion using the
solution provided by \citet{bell+11}. Then, we create a master-frame list using
one of the {\em F814W} exposures and we transform and average the exposures
for the other filters into the reference system defined by the master frame by
means of a six-parameter linear transformation. To obtain the final catalog we
use the FORTRAN software package KS2 (J.\ Anderson, in preparation). This
program allows us to find and measure stars in all the individual exposures
simultaneously and in particular to measure faint objects that cannot be
detected in individual exposures. Photometry of saturated stars is measured
following the method developed by \citet{gill04} for the ACS camera and
subsequently demonstrated to be valid also for the WFC3/UVIS detector
\citep{gill+10}, which allows one to recover the electrons that have bled into
neighbouring pixels.  

To obtain the photometric calibration, we compare PSF-based instrumental
magnitudes measured on the {\it flc} exposures with aperture-photometry
magnitudes measured on the {\it drc} exposures (i.e., the distortion-corrected
and resampled {\it flc} files) using a fixed aperture of 10 pixels. We then
transform the magnitudes into the VEGAMAG system by adopting the relevant
synthetic
zero-points\footnote[3]{http://www.stsci.edu/hst/wfc3/analysis/uvis\_zpts}.
Finally, for our analysis we select only those sources that are recovered in
both exposures for the filters {\em F275W}, {\em F467M}, and {\em F814W}, and in
two out of three exposures for the filter {\em F656N}, and we clean the obtained
catalog using the quality-of-fit (QFIT) parameter, provided by KS2, which
identifies how well a source has been fit with the PSF model \citep{ande+08}.  

Completeness as well as the photometric-error distribution of the final
photometry are characterized by performing artificial star tests. We use the
standard technique of adding artificial stars to the images and running them
through the photometric routines applied in the reduction process. The method
adopted to perform the artificial star tests is described in \citet{ande+08}.
Briefly, we generate a list of $3 \times 10^5$ artificial stars and we
distribute them in magnitude in order to reproduce a luminosity function similar
to the observed one and with a colour distribution that span the full colour
ranges found in the CMDs. The overall distribution of the inserted artificial
stars follows that of the stars in the image. Each star in the list is inserted
one at a time in the images with the appropriate flux and position and it is
measured using the same procedure and PSF adopted for real stars. We consider the inserted stars recovered if the input and output magnitudes agree to
within 0.75 mag in every filter and if the input and output position agree to
within 0.5 pixel. Finally, a completeness fraction is assigned to each individual star in a given CMD as a function of its magnitude and distance from the cluster center.  

The left panel of Fig.\,\ref{f:fov} shows a trichromatic stacked image of the
WFC3/UVIS field of view (red channel: {\em F814W} filter; green channel: {\em
F656N} filter; blue channel: {\em F275W} filter). The white circle has a radius
of $\approx$ 20\arcsec\ which corresponds to the effective radius $r_e$ of the
cluster, as derived in Sect.\,\ref{s:king}. On the left side of NGC\,1850, the
much younger cluster NGC\,1850B is clearly visible. The stacked image nicely
outlines the gas filaments which cover a large portion of the UVIS field of
view. However, as derived in Sect.\,\ref{s:reddening}, these filaments seem to
lie behind the cluster and therefore do not affect our analysis. In the
following, we define the region enclosed within a radius of $2 \times r_e$ as
``the cluster field''. In order to minimize the contamination from the stars
belonging to NGC\,1850B, we excluded from the final ``NGC\,1850 field'' catalog
objects in the annular segment in the direction of NGC\,1850B (i.e., the stars
that have a position angle  between 138\degr\ and 217\degr\ in the detector
frame and for which the distance from the cluster center is  larger than $r_e$).
In the right panel of Fig.\,\ref{f:fov}, we show the spatial positions of the
selected stars in the field of view. The objects that satisfy the aforementioned
criteria and represent the NGC\,1850 cluster field are reported as red dots.  

\section{colour-Magnitude Diagram Analysis}
\subsection{A wide MSTO region and a MS split}
\label{s:cmd}
\begin{figure}
\includegraphics[width=1\columnwidth]{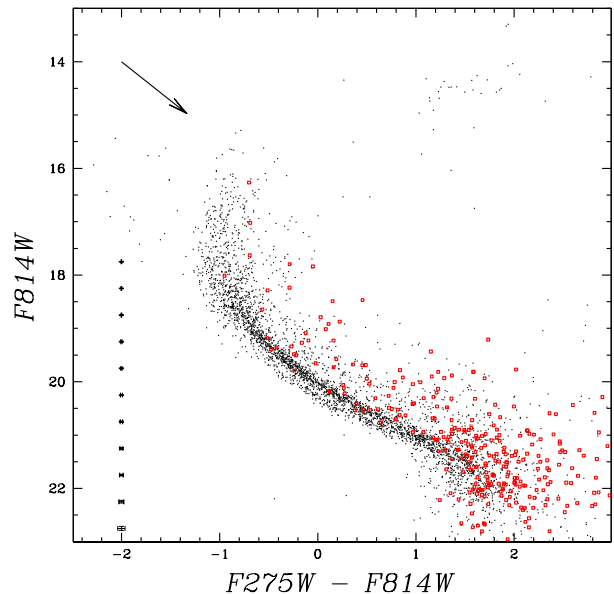}
\caption{Observed {\em F814W} vs.\ {\em F275W\,--\,F814W} CMD for all the stars
inside the NGC\,1850 field, defined in Sect.\,\ref{s:data} and shown in the
right panel of Fig.\,\ref{f:fov}. We determine the contamination from the
underlying LMC field population from a region, with the same surface area
adopted for the cluster star, near the top-right corner of the image. Stars
located in this region are superposed on the cluster CMD (red squares).
Magnitude and colour errors are shown in the left side of the CMD. The reddening
vector is also shown for $A_V = 0.5$.}
\label{f:cmdobs}
\end{figure}

Fig.\,\ref{f:cmdobs} shows the {\em F814W} vs.\ {\em F275W\,--\,F814W} CMD of
the NGC\,1850 field, derived as described in Sect.\,\ref{s:data} and reported in
the right panel of Fig.\,\ref{f:fov}. The observed CMD presents two interesting
features: (1) a MSTO region that is wider than what is expected from a single
stellar population and (2) the presence of two MSes: a blue and poorly-populated
MS, and a red MS, which contains the majority of the stars.  The two sequences
run parallel from {\em F814W} $\sim$ 18.5 mag to {\em F814W} $\sim$ 21.25 mag,
where they seem to merge together. The presence of a small colour difference
between the two sequences at fainter luminosity, not detectable due to
photometric errors, can however not be excluded.  We calculate the ratio of
stars in the blue and the red MSes by carefully selecting stars in the two MSes
in a fixed magnitude interval (19.5 mag $<$ {\em F814W}\, $<$ 21.0). We derive
that the less-populated blue MS contains around 25\% of stars, with the
remaining 75\% belonging to the red MS (see Sect.\,\ref{s:radial} for a detailed
discussion). This ratio is very similar to the one found by \citet{milo+15a} for
NGC\,1755.  

To verify that the observed broadening of the MSTO and the double MS are
``real'' features in the cluster CMD, we checked for the following potential causes: contamination by field stars, photometric errors, and differential reddening.  

We assess the level of contamination by the field population by selecting a region near the top-right corner of the image which has the same surface area as the cluster field. The stars in this region are shown in Fig.\,\ref{f:cmdobs} as red open squares. Note that these stars mainly contaminate the lower MS, at fainter magnitudes than the observed splitting, and they do not affect the MSTO at all. We therefore conclude that field stars are not the cause of the observed split in the MS and the broadening in the MSTO region.

Photometric uncertainties are derived during the reduction process.
Magnitude and colour errors are shown in the left panel of Fig.\,\ref{f:cmdobs};
photometric errors at the MSTO level are $\simeq$ 0.005 mag, whereas they are of
the order of 0.015 mag at a magnitude {\em F814W} $\sim$ 21.5 mag, where the
MSes are clearly separated. These errors are far too small to account for the observed broadening of the MSTO. 

\begin{figure*}
\centerline{
\includegraphics[height=8.2cm]{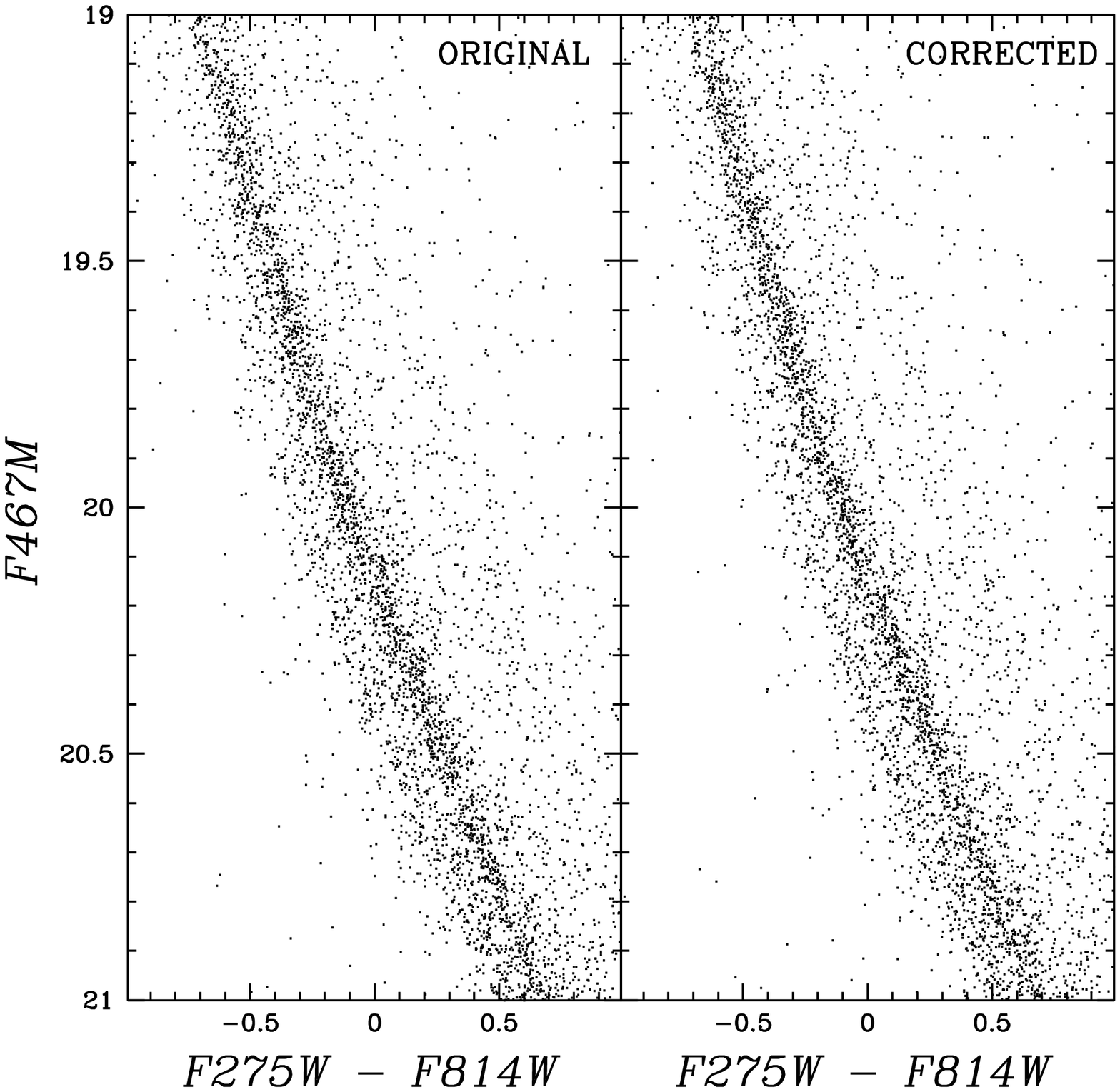}
\includegraphics[height=8.2cm]{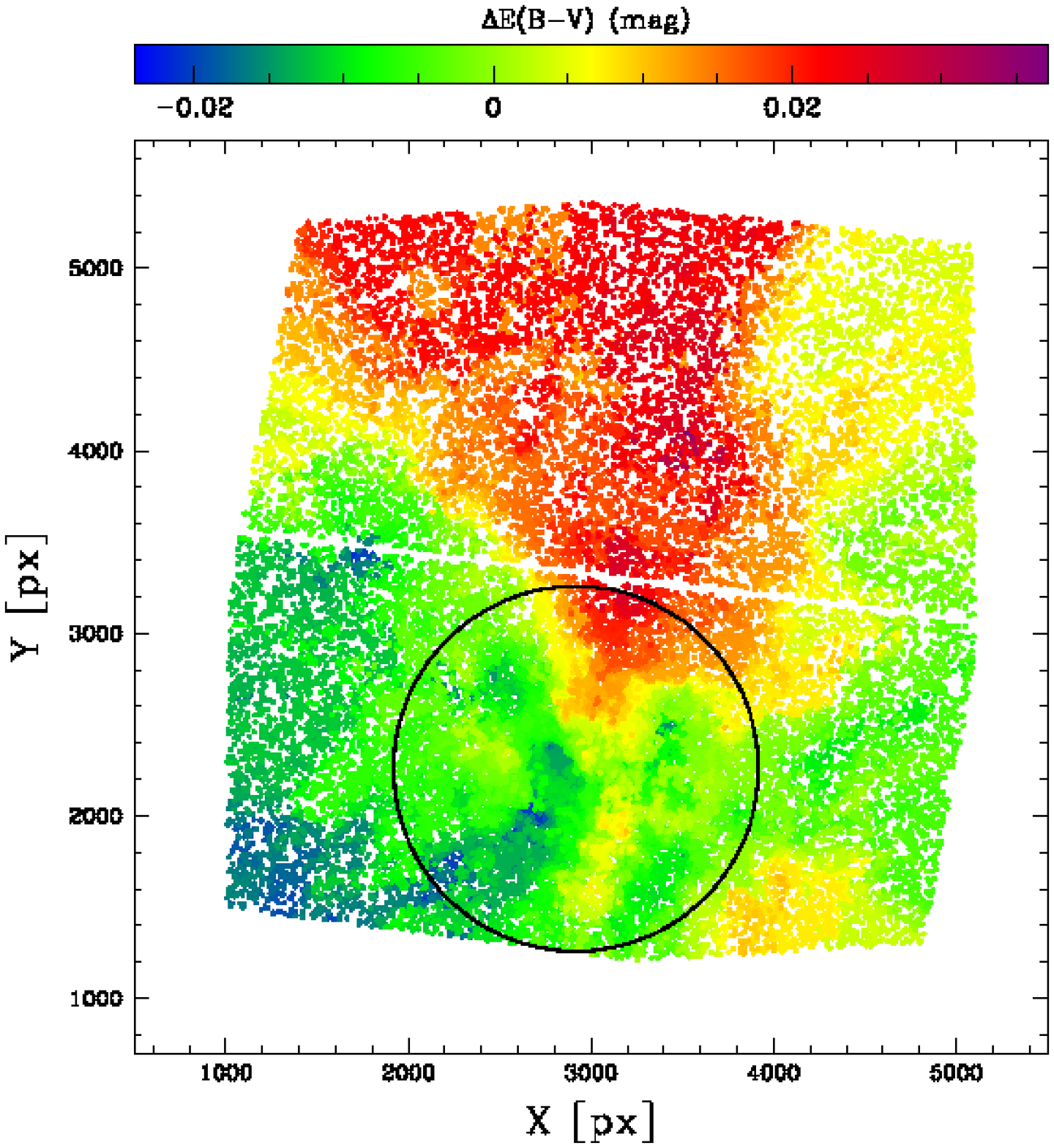}
}
\caption{Left panels: Original {\em F467M} vs.\ {\em F275W\,--\,F814W} CMD of
stars in the NGC\,1850 field of view, and CMD corrected for differential
reddening, both zoomed in the central part of the MS. Right panel: Spatial
distribution of the differential reddening in the WFC3/UVIS field of view. The
colour of each star represents the final $\Delta E(B-V)$ derived from the
differential reddening correction. The colour coding is shown at the top. The
black circle marks a diameter of $\approx$ 40\arcsec\ (i.e. $2 \times r_e$).}  
\label{f:redmap}
\end{figure*}

\subsection{The impact of differential reddening}
\label{s:reddening}
To check whether differential reddening is present in our WFC3/UVIS field of
view and, in particular, in the cluster region, we adopt the same approach as
in \citet{corr+15} and described in detail in \citet{milo+12}. Briefly, we first
rotate the CMD to a photometric reference frame in which the X-axis is parallel
to the reddening line. This reference system makes it easier to estimate
reddening differences than in the original colour-magnitude plane, where the
reddening vector is inclined. Then, we generate a fiducial line along the
most-populated red MS and, among the stars in the red MS, we select a sample of
reference stars with high-precision photometry. For each star in the field of
view, we select a sample of 50 neighbour reference stars and we calculate the
residual colours with respect to the fiducial line. The median of these
residuals, calculated along the reddening direction, is assumed as the
differential reddening corresponding to each star. Each reference star is
excluded in the computation of its own differential reddening.  

The results of the differential reddening correction are shown in
Fig.\,\ref{f:redmap}, where we compare a zoom-in portion of the MS for the
original CMD (left panel) of NGC\,1850 and the CMD corrected for differential
reddening (middle panel). The two CMDs are almost identical, suggesting that the
impact of differential reddening is negligible and that the MS split is real. As
a further proof, we show in the right panel of Fig.\,\ref{f:redmap} the spatial
distribution of the differential reddening in our field of view. Each star is
reported with a different colour, depending on the final $\Delta E(B-V)$ that we
applied. In order to better visualize how the differential reddening impacts the
area of the cluster field, we report in Fig.\,\ref{f:redmap} a circular region,
with a radius $r = 2 \times r_e$. Fig.\,\ref{f:redmap} clearly shows that the
reddening variations in the NGC\,1850 field are very small, of the order of $\pm
0.02$ mag in $E(B-V)$. The technique that we applied is able to minimize
the amount of differential reddening in the foreground of the cluster. We find a
very small reddening variation across the field of view, and with a very
specific spatial pattern that is completely uncorrelated to the \halpha\,
filaments we see in Fig.\,\ref{f:fov}. This implies that these filaments are located behind the cluster. Finally, since the reddening variation is
negligible, we use the original photometry in the analysis for simplicity. 

\section{Isochrone fitting}
\label{s:iso}
To understand the nature of the double MS and extended MSTO in NGC\,1850, we
compare the CMD with model isochrones. We  investigate the possibility that
these features are caused by the two main scenarios presented in the literature:
the presence of an age spread in the cluster or of a range of rotation
velocities among the cluster stars.   

Isochrone fitting is performed following the method described in detail in
\citet{goud+11b}, and adopted also in C15, for the case of star clusters that do not yet have a developed RGB. 

In the following Sections, we show the results of the isochrone fitting for the
two aforementioned scenarios. 

\subsection{Isochrones fitting: age spread}
\label{s:isoage}
\begin{figure}
\includegraphics[width=1\columnwidth]{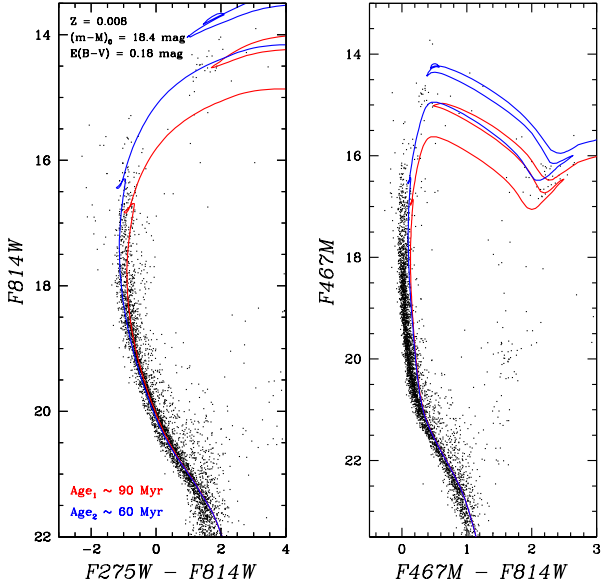}
\caption{Left panel: {\em F814W} vs.\ {\em F275W\,--\,F814W} CMD for NGC\,1850.
Best-fitting isochrones from PARSEC \citep{bres+12}, for the minimum (blue line)
and maximum (red line) age ($\sim$ 60 and $\sim$ 90 Myr, respectively) that can
be accounted for by the data are superposed on the cluster CMD, along with the
derived metallicity, distance modulus $(m-M)_0$ and reddening $E(B-V)$. Right
panel: {\em F467M} vs.\ {\em F467M\,--\,F814W} CMD. Isochrones with the same
parameters as in the left panel are overplotted on the cluster CMD.}  
\label{f:cmdage1}
\end{figure}

To test whether an age spread can reproduce the observed CMD morphology, we
superpose two isochrones from PARSEC \citep{bres+12} of different age ($\sim$ 90
Myr and $\sim$ 60 Myr) on the {\em F814W} vs.\ {\em F275W\,--\,F814W} CMD shown
in the left panel of Fig.\,\ref{f:cmdage1} (red and blue lines,
respectively). The age of the isochrones are chosen to match the maximum and
minimum age that can be accounted for by the position and width of the MSTO.
Note (a) that the morphology of the MSTO is not very well reproduced by a SSP
and that an age spread of the order of 30 Myr may constitute a better fit to the
data in this region of the CMD and (b) that the observed double MS cannot be
reproduced at all using a combination of different ages. Interestingly, we note
that if we adopt the same parameters in terms of age, metallicity, distance, and
reddening and we superpose the isochrones in the {\em F467M} vs.\ {\em  
F467M\,--\,F814W} CMD (shown in the right panel of Fig.\,\ref{f:cmdage1}) we see
that the selected isochrones do not fit the CMD  MS, the isochrones being too
red with respect to the observed MS (note that due to the shorter colour
baseline, the split is not observable in this CMD). 

\begin{figure}
\includegraphics[width=1\columnwidth]{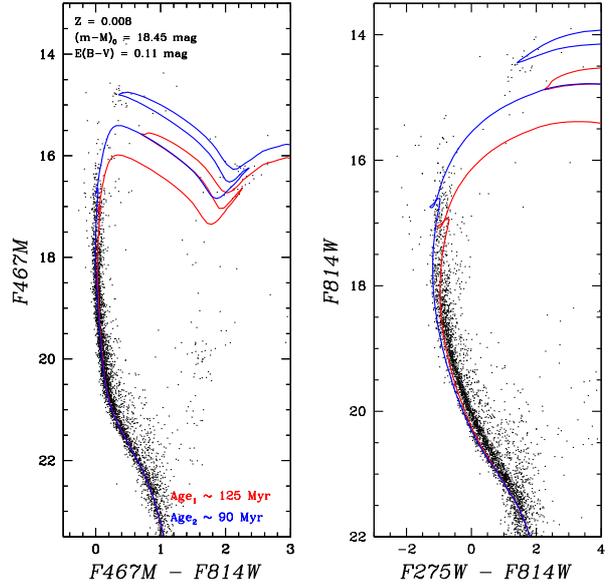}
\caption{Similar to Fig.\,\ref{f:cmdage1}. In this case, best-fitting isochrones
are derived from the {\em F467M} vs.\ {\em F467M\,--\,F814W} CMD (left panel).
The adopted age, metallicity, distance, and reddening are also reported.
Isochrones with these values are superimposed on the {\em F814W} vs.\ {\em F275W\,--\,F814W} CMD (right panel).}   
\label{f:cmdage2}
\end{figure}

To verify that this discrepancy is not due to systematic uncertainties in the
isochrone colour transformations, we also perform the inverse approach. We use
the {\em F467M} vs.\ {\em F467M\,--\,F814W} CMD (left panel of
Fig.\,\ref{f:cmdage2}) to derive the isochrone parameters and then we overplot
them onto the {\em F814W} vs.\ {\em F275W\,--\,F814W} CMD (right panel of
Fig.\,\ref{f:cmdage2}). The latter CMD shows that also   in this case a
combination of isochrones with different ages provides a satisfactory fit of the
MSTO. And, importantly, \emph{both isochrones match the less-populated blue MS
of the cluster very well}. This is a clear indication that the red MS,
containing the majority of stars, cannot be reproduced using non-rotating
isochrones with different ages and thus \emph{must be constituted by rotating
stars}. 

\subsection{Isochrones fitting: range of rotation velocities}
\label{s:isorot}
\begin{figure*}
\begin{center}
\includegraphics[height=12cm]{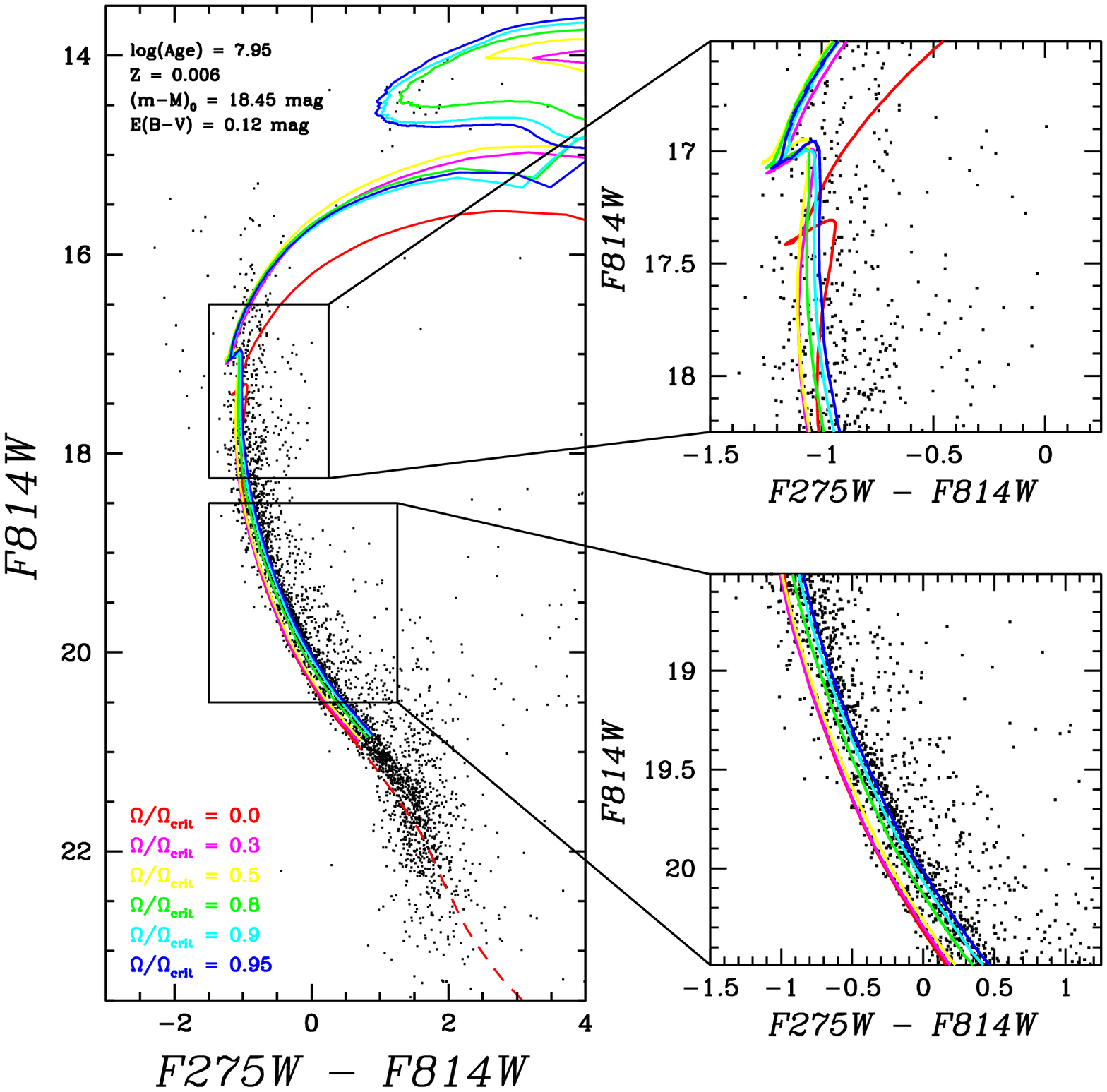}
\caption{Left panel: {\em F814W} vs.\ {\em F275W\,--\,F814W} CMD with
superimposed isochrones with different rotation rates from the Geneva database.
Isochrones are colour-coded, in terms of their \Omratio, as follows:
\Omratio\, = 0.00 (no rotation), red line; \Omratio\, = 0.30, magenta line;
\Omratio\, = 0.50, yellow line; \Omratio\, = 0.80, green line; \Omratio\, =
0.90, cyan line, and \Omratio\, = 0.95, blue line. Non-rotating isochrone, from
\citet{mowl+12}, for M $<$ 1.7 \Msun\, is shown as a dashed red line. Adopted
age, metallicity, distance, and reddening are also reported. Right panels:
zoom-in of the MSTO region (top panel) and of the MS (bottom panel) with
superimposed the same isochrones as in the left panel.} 
\label{f:cmdrot}
\end{center}
\end{figure*}

As mentioned above, the alternative scenario proposed to explain the eMSTO
phenomenon suggests that it is caused by a spread in rotation velocity among TO
stars. To test this scenario, we use the isochrones from the Geneva SYCLIST
database \citep{ekst+12,geor+13,geor+14} for which different rotation rates are
available. The left panel of Fig.\,\ref{f:cmdrot} shows the {\em F814W} vs.\
{\em F275W\,--\,F814W\,} CMD with superimposed the isochrones for different
rotation rates. In detail, the red isochrone has no rotation, i.e. \Omratio\, =
0.0, where $\Omega_{crit}$ represents the critical (break-up) rotation rate,
whereas isochrones with \Omratio\, = 0.30, 0.50, 0.80, 0.90, and 0.95 are
reported in magenta, yellow, green, cyan, and blue, respectively. The
non-rotating isochrone has been extended to masses M $<$ 1.7 \Msun\, by using
the models of \citet[][reported in the CMD as a red dashed line]{mowl+12}. Due
to slightly different input physics in the two models, a small color mismatch is
present between the two non-rotating isochrones. To account for it, we shifted
the \citet{mowl+12} non-rotating isochrone toward the blue by 0.06 mag. Rotating
isochrones with M $<$ 1.7 \Msun\, are not (yet) available. The adopted age,
distance modulus, and reddening are also reported in the left panel of
Fig.\,\ref{f:cmdrot}. For what concerns metallicity, only three options are
provided by the Geneva database for the rotating isochrones and we choose the
value most appropriate for our case (i.e., $Z = 0.006$). 

A combination of different rotation rates seems to reproduce the MS split and
the distribution of stars in the blue loop quite well. In particular,
fast-rotating stars (\Omratio\,$>$ 0.80) constitute the more populated red MS,
whereas non-rotating/slow rotating stars (\Omratio\,$<$ 0.50) constitute the
blue MS. The presence of a clear gap between the two MSes and thus the absence
of intermediate-velocity rotators (0.50 $<$\Omratio\,$<$ 0.80) indicates that
the distribution of rotation velocities is somewhat bimodal, rather than
continuous. Fig.\,\ref{f:cmdrot} also shows that the fit in the MSTO is not
satisfactory. This can be better appreciated in the right panels of
Fig.\,\ref{f:cmdrot}, in which we show  zoom-in views of the MSTO (top panel)
and the MS in the surroundings of the split (bottom panel). In particular, in
the MSTO region, while the ``spread'' in {\it F814W\/} magnitude is reproduced
by a combination of isochrones of different rotation rates, the spread in
\emph{colour} is larger in the observed CMD with respect to the one produced by
the isochrones. This is further addressed below. 

\section{Monte Carlo Simulations}
\label{s:mc_sim}
\begin{figure*}
\centerline{
\includegraphics[height=8.2cm]{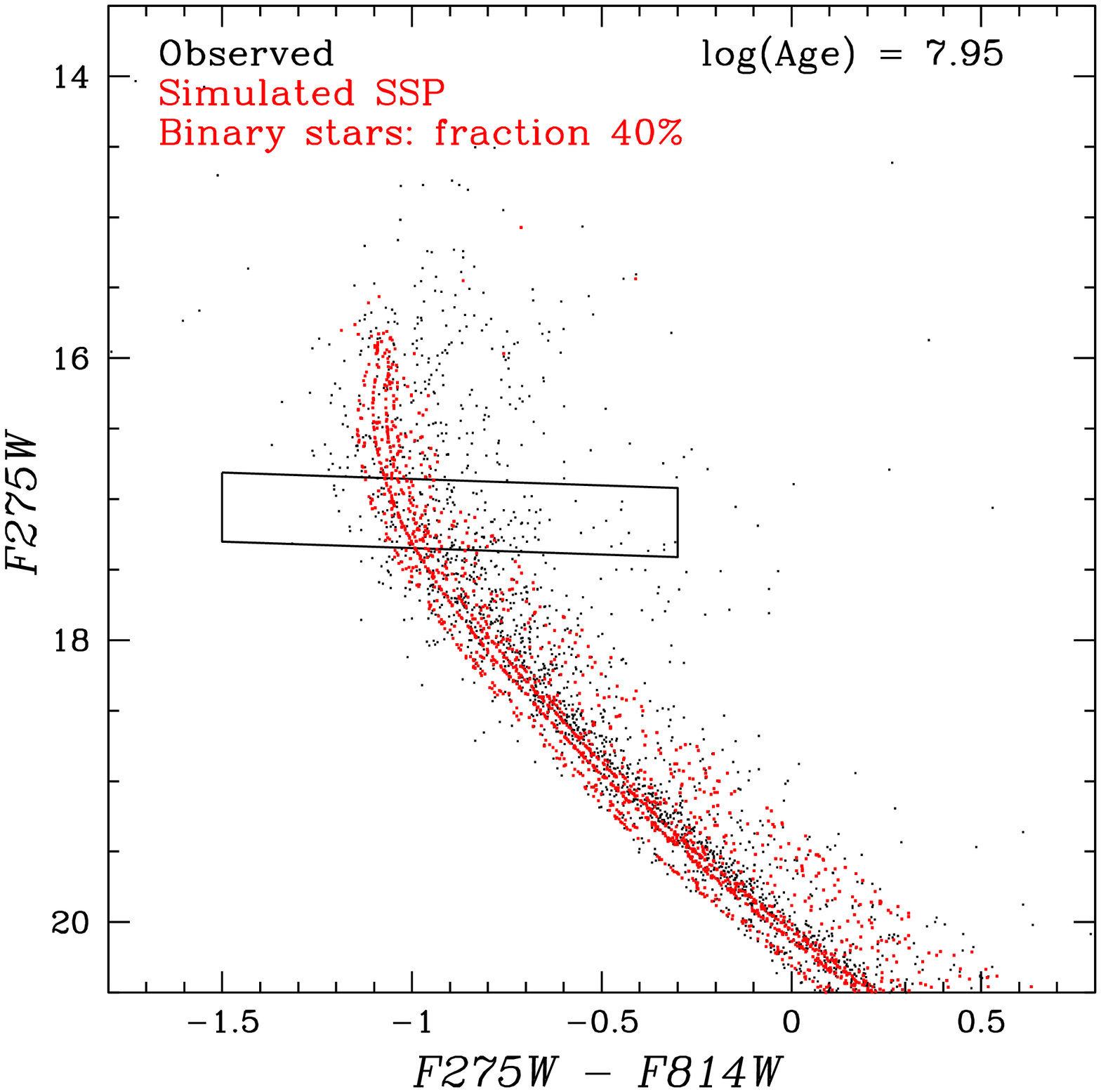}
\includegraphics[height=8.2cm]{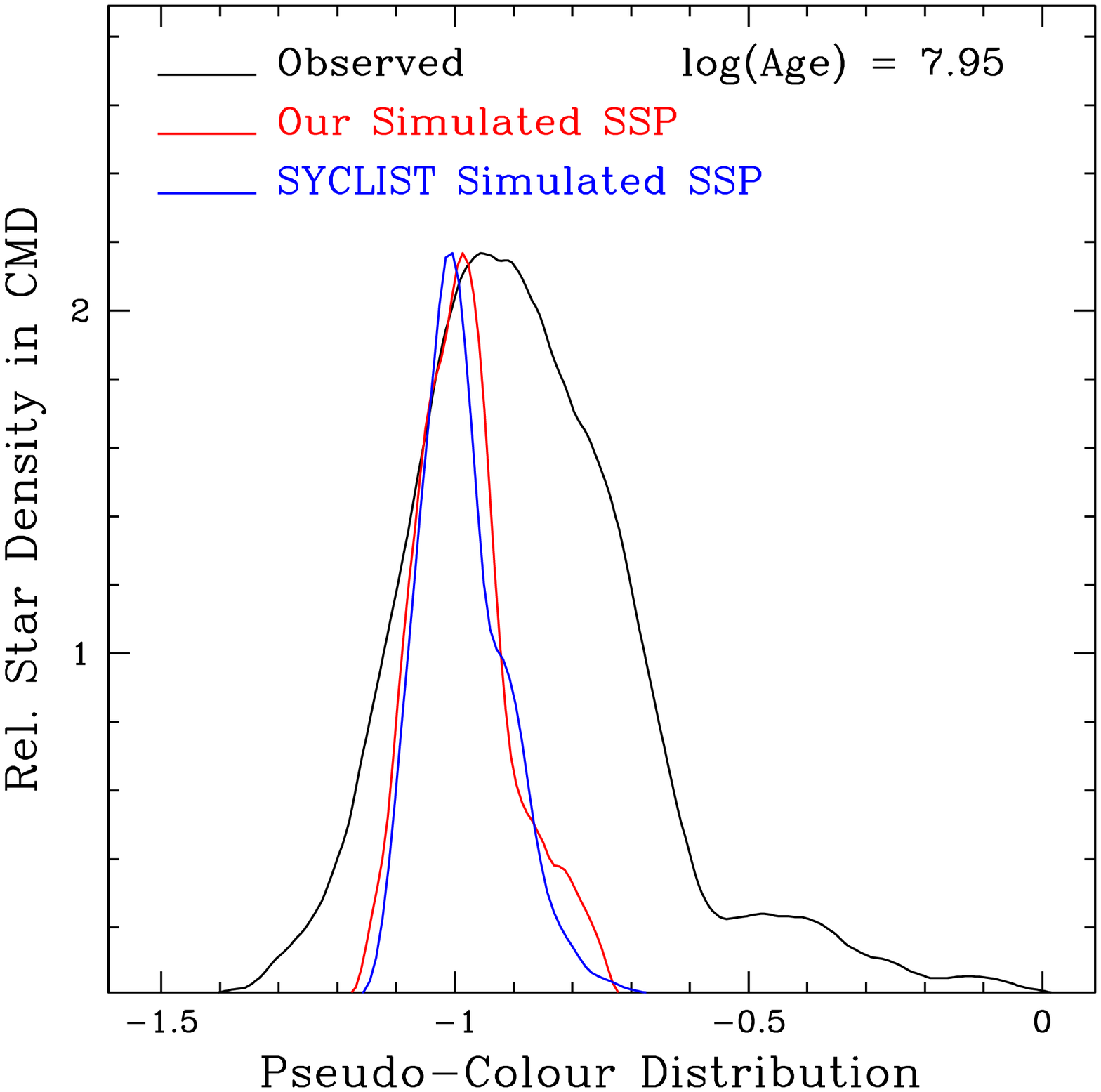}
}
\caption{Left panel: comparison between observed (black dots) and simulated (red
dots) CMDs. The latter is obtained from Monte Carlo simulations of a SSP with an
age of $\sim$ 90 Myr (log(age) = 7.95), populating non-rotating and rotating
isochrones as described in Sect.\,\ref{s:mc_sim}. We also report the
parallelogram box used to select MSTO stars. Right panel: pseudo-colour
distribution for the MSTO region of the observed (black line) and simulated (red line, our simulation; blue line, SYCLIST simulation) CMDs.}  
\label{f:cmdsim1}
\end{figure*}

To further determine to what extent a range in rotation velocities can reproduce
the observed CMD, and in particular the observed MSTO morphology, we conduct
Monte Carlo simulations of synthetic clusters with a single age and different
stellar rotation velocities \citep[the method adopted to produce these
simulations is described in detail in][]{goud+11b,goud+14}.  Briefly, we
simulate a SSP with a given age and chemical composition populating the
non-rotating and rotating isochrones. Stars are randomly drawn using a Salpeter
mass function and normalized to the observed (completeness-corrected) number of
stars. As a first guess, we arbitrarily populated the different isochrones, in
terms of \Omratio, such that the relative numbers of stars reflect the
distribution of \Omratio\ for A-type stars in the solar neighbourhood from
\citet{roye+07}. After inspection of the results, we modified the relative
numbers of stars so as to better reproduce the observed CMD morphology, that is
to match the observed ratio between the blue and the red MS (i.e., we increased
the relative numbers of stars for the fast-rotating populations with \Omratio\,
$\geq$ 0.80).   The adopted final values are the following: $\simeq$ 10\% for
\Omratio\, = 0.00, $\simeq$ 5\% for \Omratio\, = 0.30, $\simeq$ 10\% for
\Omratio\, = 0.50, $\simeq$ 35\% for \Omratio\, = 0.80, $\simeq$ 30\% for
\Omratio\, = 0.90, and $\simeq$ 10\% for \Omratio\, = 0.95.  A component of
unresolved binary stars, derived from the same mass function, is added to a
fraction of the sample stars. We use a flat distribution of primary-to-secondary
mass ratios and we adopt a binary fraction of $\simeq$ 40\%. For the purposes of
this work, the results do not change significantly within $\pm$ 20\% of the
binary fraction. Finally, we add photometric errors with a distribution derived
during the photometric reduction process. We also obtained SYCLIST cluster
simulations from their website for comparison purposes. Simulations include
gravity and limb darkening effects \citep{esplar11,clar00} as well as a random
distribution of inclination angles \citep[rotation velocity distribution
from][]{huang+10}. We found that the MSTO morphologies of the two sets of
simulations are fully consistent with one another (see Sect.\,\ref{s:pseudo}
below and right panel of Fig.\,\ref{f:cmdsim1}). We derive four different sets
of synthetic CMDs, with the only difference being the adopted age. In
particular, we simulate four SSPs with log(age) values of 7.95, 8.00, 8.05, and
8.10.   

\begin{figure}
\begin{center}
\includegraphics[width=1\columnwidth]{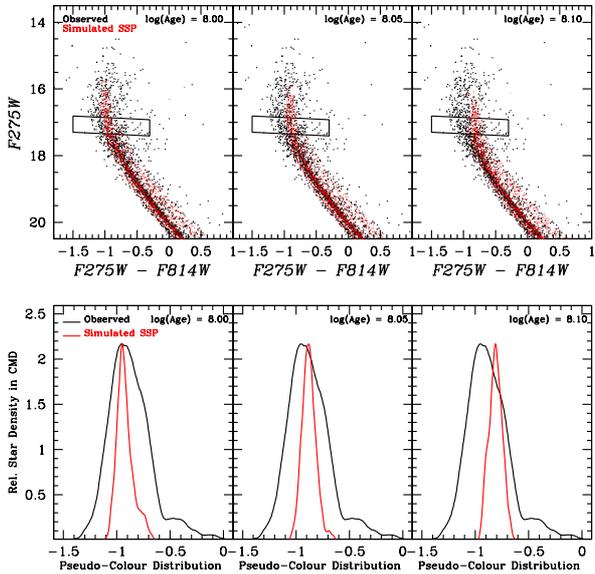}
\caption{Top panels: comparison between the observed and simulated CMDs,
obtained from SSP with ages log(age) = 8.00 (left panel), log(age) = 8.05
(middle panel), and log(age) = 8.10 (right panel). Bottom panels: correspondent
pseudo-colour distributions for the MSTO regions.} 
\label{f:cmdsim2}
\end{center}
\end{figure}

To compare in detail the observed MSTO region with the simulated ones, we create
pseudo-colour distributions. Briefly, we construct a parallelogram across the
MSTO (shown in the left panel of Fig.\,\ref{f:cmdsim1}) and we select the stars
within it. We define the distributions with the term ``pseudo-colour'' because
it reflects the star distribution along the major axis of the parallelogram,
rather than the {\em F275W\,--\,F814W\/} colour. Then, we calculate the 
pseudo-colour distributions using the non-parametric Epanechnikov-kernel density
function \citep{silv86}.  This is done to avoid possible biases that can arise
if fixed bin widths are used. We apply the procedure to both the observed and
the simulated CMD. The results obtained from the comparison of the observed and
simulated pseudo-colour distributions are described in the following Sections.

\subsection{Testing the influence of stellar rotation at the MSTO}
\label{s:pseudo}
The left panel of Fig.\,\ref{f:cmdsim1} compares the observed and simulated CMD,
the latter obtained as described above, adopting a single SSP with an age of
$\sim$ 90 Myr (log(age) = 7.95) and a distribution of rotation rates that fits
the morphology of the MS split. As expected from the isochrone comparison
described in the previous Section, this simulation reproduces the MS split quite
well. Conversely, the MSTO region of the simulation does not reproduce the
observed MSTO region well, the latter being wider than the former. This is
highlighted in the right panel of Fig.\,\ref{f:cmdsim1}, in which we compare the
observed and simulated pseudo-colour distributions. Indeed, the observed
distribution (shown as a black line) is wider than the simulated one (red line);
in particular, while the distributions are similar in the left half (i.e., blue
side) of their respective profiles, the observed one extends towards a
significantly redder colour than the simulated ones. The same results are
obtained comparing the observed distribution with that obtained from the SYCLIST
simulation (reported as a blue line in the right panel of
Fig.\,\ref{f:cmdsim1}). Hence, Fig.\,\ref{f:cmdsim1} suggests that the inclusion
of gravity and limb darkening effects, as well as the inclusion of a random
viewing angle, cannot account for the observed MSTO morphology, since the width
and shape of the two simulated pseudo-colour distributions are almost identical.
 
\begin{figure*}
\centerline{
\includegraphics[height=8.2cm]{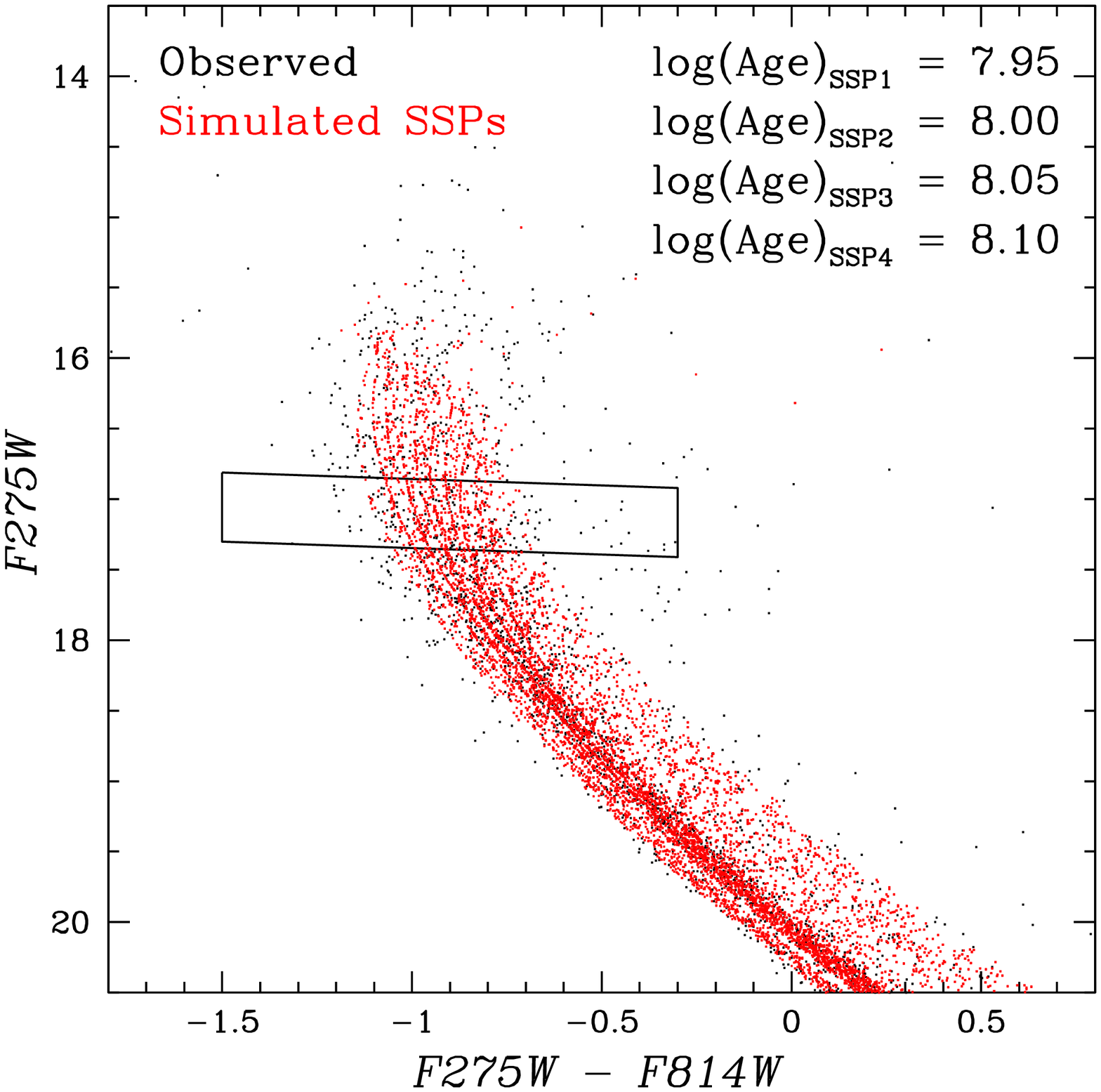}
\includegraphics[height=8.2cm]{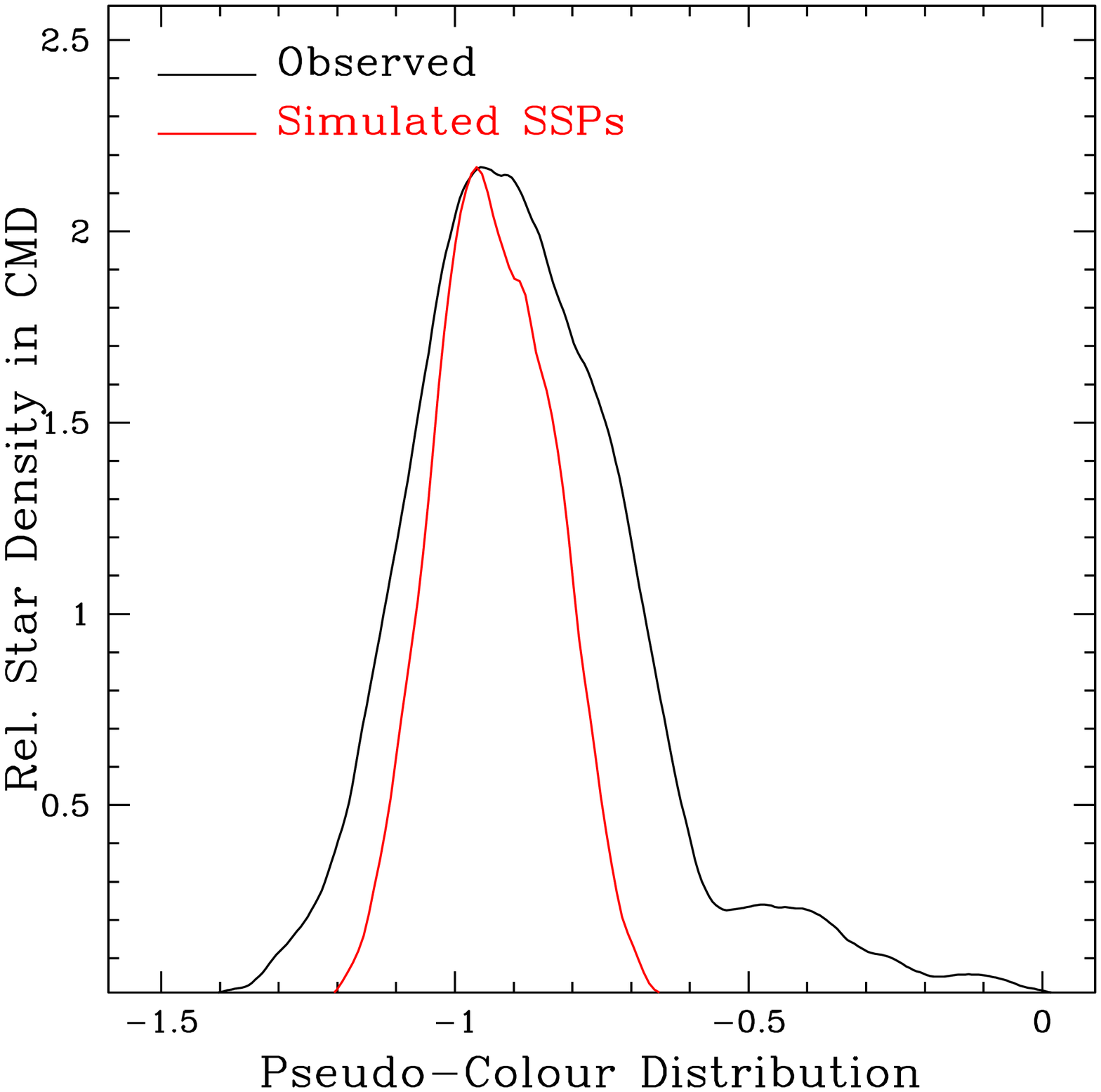}
}
\caption{Left panel: comparison between observed (black dots) and simulated (red dots) CMDs. The latter is obtained combining four SSPs with different ages ($\sim$ 90 Myr, $\sim$ 100 Myr, $\sim$ 110 Myr, and $\sim$ 125 Myr) and different rotation rates. Right panel: pseudo-colour distribution for the MSTO region of the observed (black line) and simulated (red line) CMDs.} 
\label{f:cmdsim3}
\end{figure*}

To verify that the obtained results do not depend on the adopted age, in the top
panels of Fig.\,\ref{f:cmdsim2}, we compare the observed CMD with the other
three SSP simulations which have older ages. The comparison between the observed
and simulated pseudo-colour distributions for these SSPs is shown in the bottom
panels. This comparison clearly shows that, as expected for older ages, the peak
of the simulated distribution moves towards the right (i.e., towards redder
colours), but, independently of the adopted age, the observed distribution is
always wider than the simulated ones. Thus, it seems that a range of rotation
velocities, which can reproduce the MS split, cannot explain the observed MSTO
morphology.  

\subsection{Testing the combination of stellar rotation and age spread}
\label{s:pseudo2}
In this context, we test the possibility that the eMSTO can be caused by a
combination of the two effects, that is the presence of an age spread \emph{and}
a range of stellar rotation velocities for the cluster stars.  To test this
hypothesis, we derive the combined pseudo-colour distribution of the 4 SSPs. At
this stage, we are only interested in verifying whether the combination of a
range of SSP ages provide a pseudo-colour distribution width  that can match the
observed one, and not in a detailed comparison of their global profiles. Hence,
we assume that the four SSPs have the same number of stars in the CMD, without
tailoring the relative weights based on the observed profile. The left panel of
Fig.\,\ref{f:cmdsim3} shows the comparison between the observed and simulated
CMD, whereas in the right panel we plot the corresponding pseudo-colour
distributions. The right panel clearly shows that the two distributions are
similar and that their widths are comparable (see Sect.\,\ref{s:halpha} below
for a possible explanation of the discrepancy between the two distributions at
the red end). It is also worth to note that the obtained distribution is not
obviously bimodal and is well approximated by a single Gaussian. This indicates
that the profile of the distribution itself cannot be used to exclude the
presence of an age spread, in contradiction with the claims of
\citet{bast+16}\footnote[4]{It is premature to predict exactly when secondary
star formation might have occurred after the first burst of star formation,
since the gas input can in principle come from various sources (e.g., accreted
``pristine'' gas and/or slow winds from massive stars or AGB stars).}.

Hence, the results presented in these Sections show that a single SSP with
different rotation rates fails in reproducing the observed pseudo-colour
distribution of the MSTO whereas a combination of four SSPs with different ages
and different rotation rates can provide a better fit. This suggests  that both
effects are necessary to reproduce the observed MSTO region morphology. 

\section{Insights from H${\alpha}$ analysis}
\label{s:halpha}
One of the original goals of this work was to determine whether NGC\,1850 hosts 
{\em ongoing\/} star formation. Although the results from the analysis of the
MSTO region presented in Sect.\,\ref{s:iso} and \ref{s:mc_sim} seem to exclude
such presence, a definitive answer can be provided by performing a search for
pre-main sequence stars exploiting narrow-band imaging of the \halpha\  emission
line. 

To do this, we use the method adopted in C15 and described in detail in
\citet{dema+10}. 

\begin{figure}
\includegraphics[width=1\columnwidth]{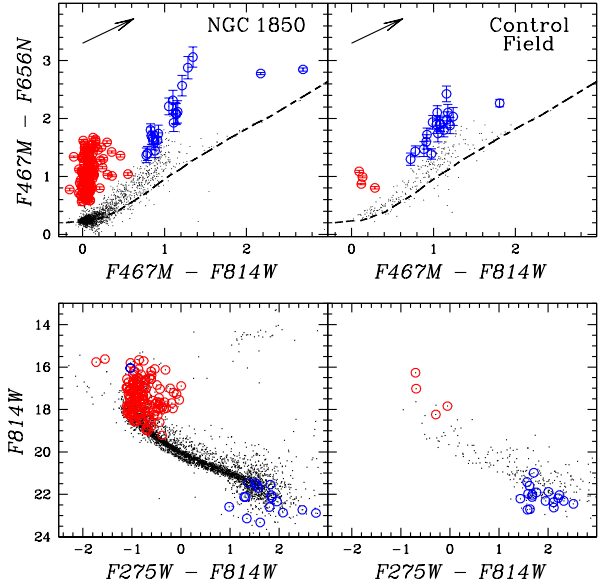}
\caption{Top panels: {\em F467M\,--\,F656N} vs.\ {\em F467M\,--\,F814W}
colour-colour diagram for the stars inside the NGC\,1850 field (left panel) and
in the control field (right panel), representing the contamination from the
background LMC stars. The dashed line represents the median {\em
F467M\,--\,F656N} colour of stars with small photometric uncertainties in each
of the three bands, representative of stars with no \halpha\  excess.
H$\alpha$-emitting stars that satisfy the criteria explained in
Sect.\,\ref{s:halpha} are shown as open red or blue circles, depending on their
{\em F467M\,--\,F814W\/} colour. Bottom panels: CMDs of the cluster (left panel)
and of the control field (right panel). Stars with \halpha\ excess are shown as
red or blue open circles as explained above.}  
\label{f:excess}
\end{figure}

The top panels of Fig.\,\ref{f:excess} show the {\em F467M\,--\,F656N\/} vs.\
{\em F467M\,--\,F814W\/} colour-colour diagram for the stars in the cluster
field (top-left panel) and for the stars in the ``control field'' (top-right
panel), the same field we used to derive the background contamination in
Fig.\,\ref{f:cmdobs}. Using the median {\em F467M\,--\,F656N\/} colour of stars
with small photometric uncertainties in each of the three bands as a function of
{\em F467M\,--\,F814W}, we define the reference sequence with respect to which
\halpha\ emission excess is identified (shown as the dashed line in
Fig.\,\ref{f:excess}). To select a first candidate sample of stars with
\halpha\  excess emission, we consider all those objects with a  {\em
F467M\,--\,F656N\/} colour at least 5$\sigma$ above the reference sequence,
where $\sigma$ is the uncertainty of  the {\em F467M\,--\,F656N\/} colour for
each star. Then, we calculate the equivalent width of the \halpha\ emission line
($EW_{H{\alpha}}$) from the measured colour excess, using the following equation
from \citet{dema+10}: 
\begin{equation}
EW_{H{\alpha}} = RW \times \left[1 - 10^{-0.4 \times (H{\alpha}-H{\alpha}^c)}\right]
\label{eq:ha_ex}
\end{equation}
where RW is the rectangular width of the filter, similar in definition to the
equivalent width of the line \citep[for the adopted {\em F656N} filter, $RW$ =
27.61; see Table\,4 in][]{dema+10} and \halpha\, - $H{\alpha}^c$ is the
difference between the measured \halpha\ magnitude and the \halpha\ continuum,
represented in this case by the difference between the star {\em
F467M\,--\,F656N\/} colour and the reference sequence {\em F467\,--\,F656N\/}
colour, at the corresponding {\em F467M\,--\,F814W}. Finally, we keep only the
objects for which $EW_{H{\alpha}} > 10$ \AA\  \citep{whibas03}. The stars that
satisfy these criteria are reported as red or blue open circles in
Fig.\,\ref{f:excess}, depending on their {\em F467M\,--\,F814W\/} colour.  

The top panels of Fig.\,\ref{f:excess} show that for {\em F467M\,--\,F814W\/}
$\ga$ 0.7 mag, the number of stars with \halpha\ excess is similar in both
fields (blue open circles). However, there is a large number of objects in the
NGC\,1850 field with {\em F467M\,--\,F814W\/} $\simeq$ 0.0\,--\,0.2 mag (see red
open circles), for which there is almost no counterpart in the control field
(only four objects). In the lower panels of  Fig.\,\ref{f:excess}, we
superimpose these objects on the {\em F814W} vs.\ {\em F275W\,--\,F814W\/} CMDs.
Note that they are located in the MSTO region, covering its whole extension. In
particular, we derive that in the parallelogram box used to select MSTO stars in
the previous Sections, they represent almost the 40\% of the total number of
stars. The presence of  \halpha\ excess in stars with absolute magnitude
$M_{F814W} \sim -1$ suggests that these objects are Be stars. The emission lines
from these stars arise due to ionized circumstellar material ejected along the
poles of rapidly rotating B stars at $\Omega / \Omega_{crit} \ga 0.5$
\citep[e.g.,][]{stemei09}. Pending spectroscopic confirmation of these objects
as Be stars, these results constitute direct evidence that NGC\,1850 hosts a
significant number of fast-rotating stars.  In particular, we note that around
$-0.5 \la$ {\em F275W\,--\,F814W} $\la -0.1$ and {\em F814W} $\sim$ 18 mag,
there is a group of stars with \halpha\ excess, which are clearly redder than
the majority of stars populating the MSTO region. We suggest that this may be
due to the presence of a dusty circumstellar disk that is oriented in a close to
edge-on configuration, thus providing a ``local'' source of reddening that
cannot be corrected using the differential dereddening technique adopted in this
work.    

\begin{figure}
\includegraphics[width=1\columnwidth]{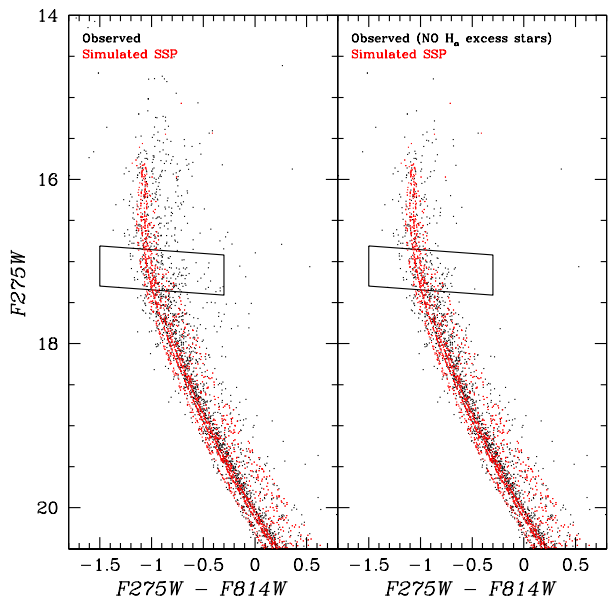} 
\caption{Comparison between the observed CMD (left panel) and the one obtained
after removing all stars that show an \halpha\ excess (right panel). In both
panels we overplot (red dots) the CMD obtained from the Monte Carlo simulations
of a SSP (the same as in Fig.\,\ref{f:cmdsim1}). The parallelogram used to
obtain the pseudo-colour distribution is also shown.}  
\label{f:cmdHa}
\end{figure}

\begin{figure}
\includegraphics[width=1\columnwidth]{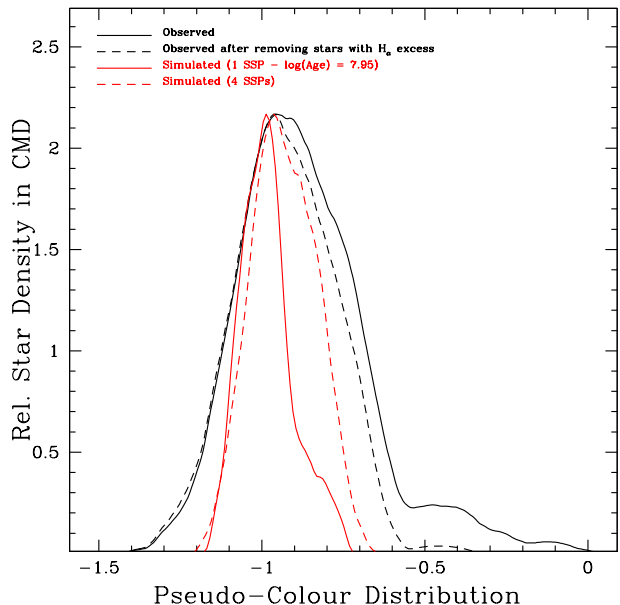}
\caption{Comparison between the observed pseudo-colour distribution obtained
from all the stars in the MSTO region (black line), the one obtained after
removing from the sample the stars with \halpha\ excess (dashed black line), the
one derived from a SSP (red line), and the one derived from the combination of
four SSPs (dashed red line).}  
\label{f:psHa}
\end{figure}

To test whether these objects may be responsible for the discrepancy between the
observed pseudo-colour distribution and the simulated one (cf.\ Sect.\,\ref{s:pseudo2}), we adopt the following empirical approach. We remove
all stars that show an \halpha\ excess (independently of the derived equivalent
widths)  from our NGC\,1850 catalog. In  Fig.\,\ref{f:cmdHa} we show the
comparison between the observed CMD (left panel) and the one obtained after the
stars with \halpha\ excess are removed from our sample (right panel). For
comparison purposes, we also superimpose onto both CMDs the same simulated
single-age CMD as in Fig.\,\ref{f:cmdsim1} (shown as red dots). As
Fig.\,\ref{f:cmdHa} clearly shows, the removal of H$\alpha$-emitting stars
results in the removal of almost all stars that were significantly redder than,
and detached from, the MSTO. Then, we derive the pseudo-colour distribution for
this subset of stars and compare it with the original one. This comparison is
shown in Fig.\,\ref{f:psHa}. Note that the red tail of the original distribution
with pseudo-colour $\ga -0.5$ mag is absent  in the new one, which is also
somewhat narrower than the original one. However, the comparison with the
simulated single-age pseudo-colour distribution (red line in Fig.\,\ref{f:psHa})
shows that the latter is still significantly narrower than the observed
distribution. We also show in Fig.\,\ref{f:psHa} the pseudo-colour distribution
obtained from the combination of four simulated SSPs (represented by the red
dashed line). In this case, the simulated distribution provides a good match to
the observed one, better in fact than the original observed one that included
the Be stars (cf.\ Sect.\,\ref{s:pseudo2}). However, we note that a small
discrepancy still persists at the red end of the distribution. This seems to
indicate that an even older population can be present in the cluster and that
the derived age spread of $\sim$ 35 Myr can be considered as a lower limit.     

These results strongly suggest that the population of \halpha-emitting stars in
the MSTO region cannot account alone for the difference between the observed
pseudo-colour distributions and those of the single-age simulations, confirming
that rotation alone cannot produce the observed morphology of the eMSTO.  

\section{Insights from the radial distribution of the two Ms populations}
\label{s:radial}
To investigate whether the population ratio between the two MSes (i.e., between non-rotating or slow-rotating and fast-rotating stars) changes as a function of cluster-centric radius we use the following method. First, we compute a fiducial line for the blue and red MS and we derive the verticalized CMD in such a way that the blue and the red fiducial translate into vertical line with abscissa 0 and 1, respectively \citep{milo+15b}. To do this we define for each star $\Delta^N_{col} = [(col - col_{blue fiducial}) / (col_{red fiducial} - col_{blue fiducial})]$, where {\em col} is {\em F275W\,--\,F814W\/}. This particular CMD allows to separate the two MSes from each other and from the binary sequence more clearly. 

\begin{figure}
\includegraphics[width=1\columnwidth]{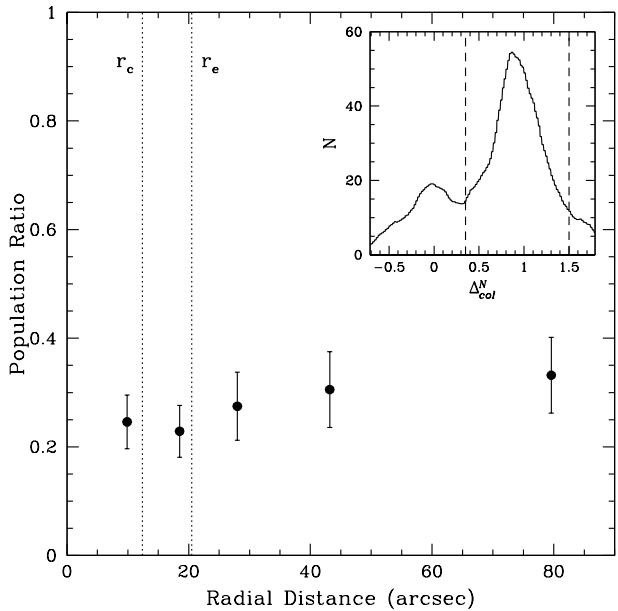}
\caption{Fraction of blue-MS stars with respect to the total number of MS stars as a function of the radial distance from the cluster center (in arcsec). The dotted vertical lines mark the projected core and effective radius. The inset panel shows the observed histogram distribution for the third radial interval. The $\Delta^N_{col}$ limits adopted to associate stars to the red and blue MS are reported as dashed lines.}  
\label{f:rad}
\end{figure}

We focus our analysis in the magnitude interval 19.5 $<$ {\em F814W} $<$ 21.0 mag. In particular, the lower magnitude limit has been chosen to exclude the faint part of the two MSes, where they seem to merge together and thus the separation is not clear. Finally, we derive the distribution of stars perpendicular to the MS in five cluster-centric annuli, chosen in such a way that they contain the same number of stars. In the inset of Fig.\,\ref{f:rad} we show an example of the observed histogram distribution for the third radial interval. We applied a smoothed na\"{i}ve estimator \citep{silv86} to be insensitive to a particular binning starting point. To determine the fraction of blue MS stars with respect to the total number of stars in each region, we adopt a $\Delta^N_{col}$ cut in the observed histogram distribution: stars with $\Delta^N_{col} <$ 0.35 mag are associated to the blue MS, whereas stars with 0.35 $< \Delta^N_{col} <$ 1.5 mag are associated to the red MS (the two limits are reported in Fig.\,\ref{f:rad} as dashed lines). The latter $\Delta^N_{col} $ limit has been chosen to avoid the contamination from binary stars.  

In Fig.\,\ref{f:rad}, we show the fraction of blue MS stars with respect to the total number of MS stars as a function of the radial distance from the cluster center. Fig.\,\ref{f:rad} shows that this ratio is constant with radius to within the Poisson uncertainties. To test the robustness of this result, we also adopt a different method to derive the fraction of blue and red MS stars in the different cluster-centric annuli. We use a bi-Gaussian function to fit the observed histogram distribution by means of least squares and we infer the fraction of stars in each MS from the areas under the Gaussians. The results obtained from the two different methods are consistent within the errors. 

Taken these results at face value, it seems that the non-rotating/slow-rotating and fast-rotating stars share the same radial distribution. Recently, similar studies have been carried out for other two young star clusters: \citet{li+16} studied NGC\,1856 and found similar results to ours for NGC\,1850, that is the two populations share the same radial distribution, whereas in the cluster NGC\,1866, \citet{milo+16b} did find a difference in the radial distribution, with the fraction of non-rotating/slow-rotating stars (blue MS) increasing slowly outward beyond its effective radius. On the other hand, \citet{goud+11b} determined the radial distribution of eMSTO stars in several intermediate-age (1\,--\,2 Gyr) massive star clusters. They find that, for the clusters with the highest escape velocities and longest relaxation times, the stars in the brightest half of the MSTO region are significantly more centrally concentrated than the stars in the faintest half. This has been interpreted in terms of an age spread; the indication that the inner region of the cluster are preferentially populated by second-generation stars is in agreement with the prediction of the {\it in situ\/} scenario \citep{derc+08}. Hence, if we assume that the results obtained here are valid also for intermediate-age star clusters (i.e., non-rotating and fast-rotating stars have the same radial distributions), it seems highly unlikely that the observed different central concentration can be explained in terms of a range of rotation velocities in the massive intermediate-age clusters. This suggests that the eMSTO phenomenon in intermediate-age star clusters may be due, at least in part, by a spread in age. Unfortunately, a direct comparison with the radial distribution of MSTO stars in NGC\,1850 is hampered by the low number of objects in the cluster MSTO region.  

\section{Insights from dynamical analysis}
\label{s:dyn}

One of the important features of the eMSTO phenomenon in intermediate-age star
clusters in the Magellanic Clouds is that they do not occur in {\em all} such
clusters. A possible explanation of this in the context of multiple populations
was provided by \citet{goud+11a}, who suggested that eMSTO are only hosted by
clusters whose escape velocities were higher than the wind velocities of
first-generation ``polluters'' stars that are thought to provide the seed
material for secondary star formation, during the era when the cluster contained
such stars (this is referred to as the ``early escape velocity threshold''
scenario). The most popular types of such polluter stars are {\it (i)\/} AGB
stars with $4 \la {\cal{M}}/M_{\odot} \la 8$ \citep{danven08}, {\it (ii)\/}
massive rotating stars \citep[``FRMS'',][]{decr+07}. and {\it (iii)\/} massive
binary stars \citep{demink+09}. Subsequently, the studies of \citet{goud+14} and
\citet{corr+14} suggested that the critical escape velocity\footnote[5]{Note
that these studies used a Salpeter IMF for mass determination.} for a star
cluster to be able to retain the material ejected by first-generation polluter
stars seem to be in the range of 12\,--\,15 \kms. 

Furthermore, the calculations of \citet{conspe11} show that massive clusters
should be able to sweep up significant fractions of their mass from the ambient
interstellar medium (ISM) at young ages if their environment is sufficiently
conducive. Under the assumptions that (1) the line-of-sight velocity dispersion
of the LMC ($\sigma \sim 20$ \kms, \citealt{vmd+02}) is a typical relative
velocity between a star cluster and the ambient ISM in the LMC, and (2) a
typical young cluster in the LMC has a half-mass radius of $\approx 5$ pc
(\citealt{macgil03}; C15), equations 5\,--\,8 of \citet{conspe11} suggest that
this can occur in the present-day LMC if the cluster is massive enough
(${\cal{M}} \ga 10^{5}$ \Msun) and the local ISM is dense enough but not so
dense that it would strip any gas accreted by the cluster by ram pressure in its
orbit (1 $\la$ (n/{\rm cm}$^3$) $\la$ 100 for a $10^5$ \Msun\ cluster).   

With this in mind, we determine the structural and dynamical parameters of the
cluster as described below. 

\subsection{Structural parameters} 
\label{s:king}
To determine the radial surface number density distribution of stars, we
follow the procedure described by \citet{goud+09} and adopted in C15. 

\begin{figure}
\includegraphics[scale=0.55]{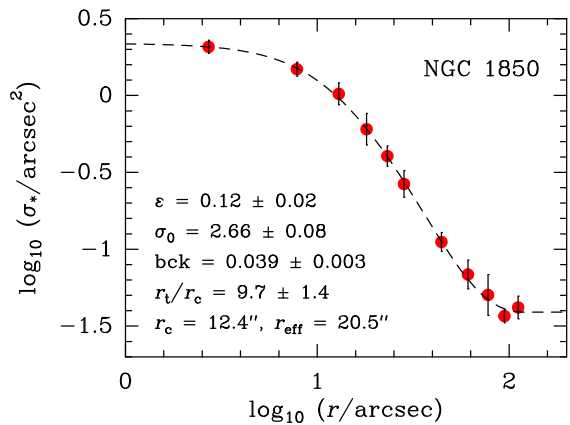}
\caption{Radial surface number density profile of NGC\,1850. Red points
represent observed values. The dashed line represents the best-fit King model
(cf. equation\,\ref{eq:King}) whose parameters are shown in the legend.
Ellipticity and  effective radius $r_e$ are also shown in the legend.}  
\label{f:king}
\end{figure}

We fit the radial surface number density profile using a \citet{king62} model
combined with a constant background level, described by the following equation: 
\begin{equation}
n(r) = n_0 \: \left( \frac{1}{\sqrt{1 + (r/r_c)^2}} - \frac{1}{\sqrt{1+c^2}}
\right)^2 \; + \; {\rm bkg} 
\label{eq:King}
\end{equation}
where $n_0$ is the central surface number density, $r_c$ is the core radius,  $c
\equiv r_t/r_c$ is the King concentration index ($r_t$ being the tidal radius),
and $r$ is the equivalent radius of the ellipse ($r = a\,\sqrt{1-\epsilon}$,
where $a$ is the semi-major axis of the ellipse and $\epsilon$ is its
ellipticity). Fig.\,\ref{f:king} shows the best-fit King model, obtained using a
$\chi^2$ minimization routine. The derived core radius $r_c$ = 12\farcs4 $\pm$
0\farcs5, which corresponds to 3.01 $\pm$ 0.12 pc, whereas the effective radius
$r_e$ = 20\farcs5 $\pm$ 1\farcs4 (4.97 $\pm$ 0.35 pc). Our estimate for $r_c$ is
in reasonable agreement with the 2.7 pc derived by \citet{nied+15}.

\subsection{Cluster mass and escape velocity}
\label{s:escvel}

\begin{table*}
\begin{center}
\begin{tabular}{cccccccccc}
\hline
\hline
\multicolumn{1}{c}{Cluster} & \multicolumn{1}{c}{I} & \multicolumn{1}{c}{Aper.} & \multicolumn{1}{c}{Aper. Corr.} & \multicolumn{1}{c}{[Z/H]} & \multicolumn{1}{c}{$A_V$} & \multicolumn{1}{c}{$r_c$} & \multicolumn{1}{c}{$r_e$} & \multicolumn{2}{c}{log (${\cal{M}}_{\rm cl}/M_{\odot}$)}\\
(1) & (2) & (3) & (4) & (5) & (6) & (7) & (8) & (9) & (10)\\
 \hline
NGC\,1850 & 9.43 $\pm$ 0.04 & 20 & 0.47 $\pm$ 0.03 & -0.3 & 0.372 & 3.01 $\pm$ 0.12 & 4.97 $\pm$ 0.35 & 4.86 $\pm$ 0.10 & 4.62 $\pm$ 0.10\\
\hline 
\end{tabular}
\caption{Physical properties of the star cluster. Columns (1): Name of the
cluster. (2): Integrated $V$ magnitude. (3): Adopted radius in arcsec for the
measure of the integrated magnitude. (4): Aperture correction in magnitude. (5):
Metallicity (dex). (6): Visual extinction $A_V$ in magnitude. (7): Core radius
$r_c$ in pc. (8): Effective radius $r_e$ in pc. (9-10): Logarithm of the cluster
mass adopting a \citet{salp55} and a \citet{chab03} initial mass function,
respectively.}
\label{t:param}
\end{center}
\end{table*} 

We estimate the cluster mass and escape velocity as a function of time going
back to an age of 10 Myr, after the cluster has survived the era of violent
relaxation and when the most massive stars of the first generation that are
proposed to be candidate polluters in literature (i.e, FRMS and massive binary
stars), are expected to start losing significant amounts of mass through slow
winds.   

The current mass of NGC\,1850 is determined from its integrated-light
\emph{I}-band magnitude listed in Table\,\ref{t:param} and derived as described
in C15. In this particular case, in order to minimize the contamination by
NGC\,1850B, we derive the aperture-photometry magnitudes of stars inside a
radius of 20$''$, centered at the coordinates found in the analysis of
Sect.\,\ref{s:king}, measured on the {\em F814W drc} image (obtained combining
long and short exposures). The background level and its uncertainty is derived
from five different regions of the image with distances larger than 100$''$ from
the center of NGC\,1850, avoiding the major \halpha\ emission filaments (see
Fig.\,\ref{f:fov}). 

We calculate the escape velocity of NGC\,1850 and its evolution with time  using
the prescriptions of \citet{goud+14}.  Briefly, we evaluate the evolution of
cluster mass and radius with and without mass segregation. This property plays a fundamental role in terms of the early evolution of the
cluster's expansion and mass-loss rate \citep[e.g.,][]{mack+08b,vesp+09}. For
the case of a model cluster with initial mass segregation, we adopt the results
of the simulation called SG-R1 in \citet{derc+08}, which involves a model
cluster that features a level of initial mass segregation of $r_e/r_{e,>1}$ =
1.5, where $r_{e,>1}$ is the effective radius of the cluster for stars with
${\cal{M}} >$ 1 \Msun\, \citep[see][for a detailed description of the reasons
for this choice]{goud+14}. Escape velocities are calculated  from the reduced
gravitational potential $V_{\rm esc} (r,t) = (2\Phi_{\rm tid} (t) - 2\Phi
(r,t))^{1/2}$, at the core radius and at the effective radius. We choose to
calculate the escape velocity at the cluster core radius in accordance with the
prediction of the {\it in situ\/} scenario \citep{derc+08}, where the second
generation stars are formed in the innermost region of the cluster. Assuming a
\citet{salp55} initial mass function, the current cluster escape velocity is
10.5 $\pm$ 0.2 \kms\ at the effective radius and 13.8 $\pm$ 0.2 \kms\, at the
core radius. For a \citet{chab03} initial mass function, we obtain 8.0 $\pm$ 0.2
\kms\, and 10.6 $\pm$ 0.2 \kms, respectively.  

\begin{figure}
\hspace*{+0.5cm}
\includegraphics[scale=0.80]{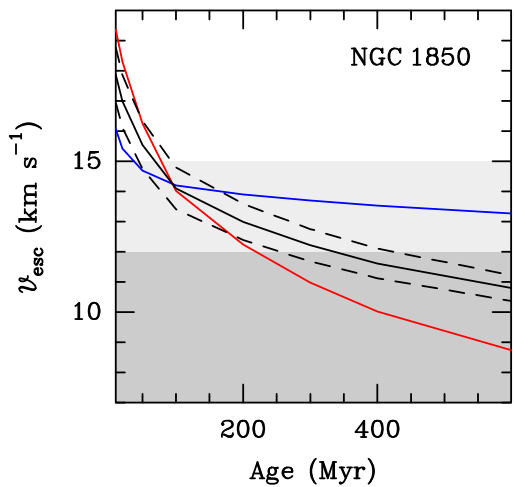}
\caption{Escape velocity as a function of time at the core radius. The black
curve represents the ``plausible'' escape velocity, derived as described in
Sect.\,\ref{s:escvel}. Formal $\pm$ 1 $\sigma$ errors of $V_{esc}$ are shown by
the dashed line. The blue line represents the escape velocity  in case NGC\,1850
had no primordial mass segregation, while the red line does so in case it had
the same level of primordial mass segregation as the model SG-R1 in
\citet{derc+08}. The light grey region represents the critical range of
$V_{esc}$ mentioned in Sect.\,\ref{s:dyn}, i.e. 12\,--\,15 \kms. The region
below 12 \kms\, in which $V_{esc}$ is thought to be too low to permit retention
of material shed by the first stellar generation, is shaded in dark grey.}   
\label{f:escvel}
\end{figure}

Fig.\,\ref{f:escvel} shows the ``plausible'' escape velocity of NGC\,1850 as a
function of time (black line), derived using the same approach as in C15 and
described in detail in \citet{goud+14}, for a \citet{salp55} initial mass
function. Briefly, the ``plausible'' escape velocity is calculated adopting a
procedure that takes into account the various results from the compilation of 
Magellanic Cloud star cluster properties and N-body simulations by
\citet{mack+08b}.

For comparison Fig.\,\ref{f:escvel} also shows the escape velocity as a function
of time for the limiting cases in which (1) the cluster does not have any
primordial mass segregation (blue line) and (2) the cluster has the same level
of primordial mass segregation as the model SG-R1 by \citet[][red
line]{derc+08}. The critical escape velocity range of 12\,--\,15 \kms\, is
depicted as the light grey region in Fig.\,\ref{f:escvel}, whereas the region
below 12 \kms, representing the velocity range in which eMSTOs are not observed
in LMC star clusters, is shown in dark grey. 

Fig.\,\ref{f:escvel} shows that, independent of the assumed level of primordial
mass segregation in the cluster, the initial escape velocity was likely high
enough to retain the material ejected through winds by the first-generation
polluters. Moreover, it seems to decrease below the critical value of $\sim$ 15
\kms\, at an age similar to what we derive as the age spread (i.e., 30\,--\,40
Myr)\footnote[7]{We emphasize that the critical value 15 \kms\ (for a Salpeter
IMF) was not more than a best-effort estimate by \citet{goud+14} based on
empirical data, and should not be interpreted as a precise number.}.   In the
context of the {\it in  situ} scenario, considering the time when
first-generation polluters are predicted to be present in the cluster (i.e., at
ages of $\sim$ 5\,--\,30 Myr for massive stars and $\sim$ 50\,--\,200 Myr for
IM-AGB stars) and the age ranges derived from our analysis (i.e., 30\,--\,40
Myr), it seems that massive stars would be the most likely source of
``polluted'' material in NGC\,1850. Future investigations of the range of
light-element abundances (e.g., Na or N) found in NGC\,1850 should be able to
shed more light on the extent to which the cluster has actually retained such
polluted material. Furthermore, the mass of NGC\,1850  is high enough for the
cluster to have been able to accrete a significant amount of ``pristine'' gas
from its surroundings in the past, which could have constituted an additional
source of gas to form stars after the initial burst that formed NGC\,1850 (cf.\
above; \citealt{conspe11}; \citealt{goud+14}; but see also \citealt{basstr14}). 

\section{Summary and Conclusions}
\label{s:summary}
We present the results of a study of new, deep {\it HST\/} WFC3 images of the
young ($\sim$ 100 Myr) star cluster NGC\,1850 in the LMC. The analysis of its
CMD reveal the presence of an eMSTO and a double MS. The blue component of the
MS hosts about a quarter of the the total number of MS stars. We verify that
these features cannot be explained in terms of photometric errors,
contamination from background LMC stars, or differential reddening.   

From a comparison with theoretical models, we show that the MS split cannot be
reproduced using non-rotating isochrones with different ages. Conversely, it
is very well reproduced by a coeval stellar population with a variety of
rotation rates. However, when adopting the latter models, we show that the
observed morphology of the MSTO cannot be fit satisfactorily: the pseudo-colour
distribution of stars across the MSTO obtained from the observed CMD is
significantly wider than that obtained from Monte Carlo simulations of the set
of single-age isochrones with a range of rotation rates that fit the double MS
feature. A combination of SSPs with an age range of $\sim 35$ Myr
\emph{along with} different rotation rates reproduces the data significantly
better. 

From an analysis of narrow-band \halpha\ imaging, we find that the MSTO region
hosts a population of Be stars, providing direct evidence that rapidly rotating
stars are present in the cluster. We note that a fraction of these stars
are redder than the majority of TO stars. We interpret this in terms of an
(close to) edge-on configuration of dusty circumstellar disks of Be stars, thus
providing a source of local reddening that cannot be accounted for in our
differential reddening correction. We quantify  how this effect impacts the
observed pseudo-colour distribution by removing all the stars with an \halpha\
excess from the original catalog and deriving a new pseudo-colour distribution.
We demonstrate that this newly derived pseudo-colour distribution is still
significantly wider than the simulated one. As such,  the eMSTO morphology still
cannot be explained by an SSP with a variety of stellar rotation rates; a spread
of ages still seems to be required. 

Finally, the dynamical properties of NGC\,1850 derived from our data show that
the cluster had an escape velocity within its core radius of $\sim$ 13\,--\,17
\kms, at an age of $\sim$ 10 Myr, high enough in principle to retain material
shed by the   slow winds of ``polluter stars''. Furthermore its mass was high
enough to accrete ``pristine'' gas that may have been present in the environment
of the cluster shortly after it was first formed. 

The important role that stellar rotation can play in the interpretation of the
eMSTO phenomenon in young and intermediate-age star clusters has been
acknowledged in the recent literature, including in works that suggested that an
age spread is responsible for eMSTOs \citep[see for example C15 and Fig.\,7
in][and discussion therein]{goud+14}. Unfortunately, the current lack of
rotating isochrone models for masses M $<$ 1.7 \Msun\, prevents us from testing
in detail its impact in intermediate-age star clusters. At the same time, the 
new findings for the young clusters do not yet provide a significant
clarification of the picture in terms of the role of stellar rotation in the
eMSTO phenomenon. For example, \citet{nied+15} concluded that the eMSTO of
intermediate-age star clusters can in principle be explained in terms of stellar
rotation, as long as the initial rotation distribution covers the interval from
\Omratio\, $\sim$ 0.0 to $\simeq$ 0.5. Conversely, the detailed studies
of the split MS of NGC\,1850 and other young clusters show that they can be
reproduced only when assuming that a large number of stars are \emph{fast}
rotators (i.e., \Omratio\, $\ga$ 0.80).  Another open issue is the origin of the
non-rotating or slow-rotating stellar populations \citep[see][for a detailed
discussion]{milo+16a}. \citet{zorroy12} suggested that the observed 
slow-rotating and non-rotating stars could have lost their angular momentum
during the pre-MS phase do to magnetic breaking. An alternative hypothesis is
that tidal interaction in binary systems could be responsible for slowing down a
fraction of stars. In this context, \citet{dant+15} suggested that binary
synchronization could be responsible for the slow-rotating population.  

Spectroscopic studies of the distribution of stellar rotation rates in young and
intermediate-age star clusters will be of fundamental importance to test the
aforementioned scenarios and address the role of the stellar rotation in the
eMSTO phenomenon. 

\section*{Acknowledgments}
Support for this project was provided by NASA through grant HST-GO-14174 from
the Space Telescope Science Institute, which is operated by the Association of 
Universities for Research in Astronomy, Inc., under NASA contract NAS5--26555.  
We made significant use of the SAO/NASA Astrophysics Data System during this
project. THP acknowledges support by the FONDECYT Regular Project Grant (n. 1161817) and the BASAL center for Astrophysics and Associated Technologies (PFB-06).


\label{lastpage}

\end{document}